\documentclass[%
twocolumn,
 amsmath,
 aps, pra,
longbibliography
]{revtex4-1}

\usepackage{color}
\usepackage{graphicx}

\renewcommand{\Re}{\mathop{\mathrm{Re}}\nolimits}           
\renewcommand{\Im}{\mathop{\mathrm{Im}}\nolimits}   
\newcommand{\sign}{\mathop{\rm sign}} 
\renewcommand{\lambdabar}{{\mkern0.75mu\mathchar '26\mkern -9.75mu\lambda}}

\begin{document}

\title{Near field versus far field in radiative heat transfer between two-dimensional metals}

\author{Jonathan L. Wise}
\email{jonathan.wise@lpmmc.cnrs.fr}
\affiliation{Universit\'e Grenoble Alpes and CNRS, LPMMC, 25 rue des Martyrs, 38042 Grenoble, France}

\author{Denis M. Basko}
\affiliation{Universit\'e Grenoble Alpes and CNRS, LPMMC, 25 rue des Martyrs, 38042 Grenoble, France}
%

\begin{abstract}
Using the standard fluctuational electrodynamics framework, we analytically calculate the radiative heat current between two thin metallic layers, separated by a vacuum gap. We analyse different contributions to the heat current (travelling or evanescent waves, transverse electric or magnetic polarization) and reveal the crucial qualitative role played by the dc conductivity of the metals as compared to the speed of light. For poorly conducting metals, the heat current may be dominated by evanescent waves even when the separation between the layers greatly exceeds the thermal photon wavelength, and the coupling is of electrostatic nature. For well-conducting metals, the evanescent contribution dominates at separations smaller than the thermal wavelength and is mainly due to magnetostatic coupling, in agreement with earlier works on bulk metals. 
\end{abstract}

\maketitle

\section{Introduction}
\label{sec:intro}

Spatially separated objects may exchange heat via electromagnetic fluctuations~\cite{Rytov1953, Polder1971, Levin1980, Loomis1994, Pendry1999}. This radiative heat transfer arises due to electric charge density and current fluctuations inside the constituting materials, and is usually described within the phenomenological framework of fluctuational electrodynamics (FED)~\cite{Rytov1953, Polder1971, Rytov1989}, for which the critical inputs are the material response functions and the system geometry. It is now well known that in the near-field limit, energy may tunnel via evanescent electromagnetic waves causing a strong enhancement of the heat transfer, as has been observed experimentally (see the reviews~\cite{Joulain2005, Volokitin2007, Song2015, Biehs2020} and references therein).

Many theoretical works have been dedicated to different material systems in the near-field regime (\cite{Joulain2005, Volokitin2007, Song2015, Biehs2020} and references therein),
in which various models for material response have been employed and different dominant channels for heat transfer identified. The common wisdom is that the evanescent modes dominate the heat transfer when the spatial separation $d\ll\lambdabar_T\equiv\hbar{c}/T$, the wavelength of photons at temperature~$T$ (here $\hbar$ and $c$ are the Planck constant and the speed of light, respectively, and we set the Boltzmann constant to unity). Indeed, for $d>\lambdabar_T$ the evanescent waves with the typical frequency $\omega\sim{T}/\hbar$ decay exponentially outside the material, while at $d\ll\lambdabar_T$ the region of the wave vectors~$k$ occupied by evanescent waves, $k\sim1/d$, is larger than that of travelling states, $k\sim1/\lambdabar_T$~\cite{Volokitin2007}.
Importance of magnetic coupling in the near-field heat transfer between well-conducting metals has been emphasised~\cite{Chapuis2008, Chapuis2008-1}.
In the extreme near-field limit, heat transfer due to the electrostatic Coulomb interaction has also been studied \cite{Prunnila2013, Mahan2017, Zhang2018, Wang2018, Kamenev2018, Wise2020, Ying2020}.

Here we revisit this old problem, focusing on the two-dimensional (2D) geometry, and study the radiative heat current between two thin metallic sheets in vacuum within the standard FED framework.
We find two qualitatively different types of behaviour, depending on the value of the two-dimensional dc conductivity $\sigma_\text{2D}$ of the sheets. For poor conductors characterised by the condition $\mathcal{G} \equiv 2\pi\sigma_\text{2D}/c \ll 1$ (we use CGS units throughout the paper, in SI units $\mathcal{G}=(\sigma_\text{2D}/2)\sqrt{\mu_0/\varepsilon_0}$), the heat transfer turns out to be dominated by the evanescent modes at distances~$d$ extending well beyond $\lambdabar_T$, and the main coupling mechanism in the near field is electrostatic (Coulomb interaction between electrons in the two layers). For $\mathcal{G}\gg1$, the conventional situation is recovered: the crossover from near to far field occurs at $d\sim\lambdabar_T$ at not too high temperatures, and in a wide range of parameters the near-field transfer is dominated by magnetostatic (inductive) coupling between currents in the layers.

The parameter $\mathcal{G}$ characterises the impedance mismatch between a 2D metal and vacuum; its importance is not restricted to the heat transfer problem and is rather general. Notably, two distinct regimes in the behaviour of 2D plasmon polaritons for $\mathcal{G}<1$ and $\mathcal{G}>1$ have been identified \cite{Govorov1989, Falko1989, Volkov2014, Muravev2015, Gusikhin2018, Oriekhov2020}. In our heat transfer problem, we find no sharp distinction between $\mathcal{G}<1$ and $\mathcal{G}>1$, but rather a smooth crossover between the two limiting situations.

For the two-dimensional geometry considered here, it is important to realise that the heat transferred from one sheet to the other is different from the heat transferred between the two half-spaces behind the sheets (which typically include dielectric substrates). The reason is that (i)~some part of the radiation emitted by each sheet may be transmitted by the other sheet and escape to infinity or be absorbed by the substrate (even if its absorption is infinitesimal, but the substrate is thick enough), and (ii)~the substrate may emit its own radiation. Which quantity is relevant, depends on the specific experimental setup, how the temperature difference is maintained, and how the heat current is measured. In this paper, we focus on the transfer between the metallic sheets, not the half-spaces. The difference between the two quantities becomes important for the far-field contribution at $\mathcal{G}\ll1$. In particular, our result about evanescent mode dominance beyond $\lambdabar_T$ applies only to the heat current from one sheet to the other.

We also emphasize that our study applies to metals only. Optical response of Drude metals and dielectrics is governed by qualitatively different physical mechanisms: conduction electrons and optical phonons, respectively, whose response is concentrated at low and high frequencies (e.~g., the optical phonon frequency in SiO$_2$ is more than three times higher than the room temperature). We do not include the contribution of such high-frequency resonances in our model. This is a valid approximation even for bad metals at sufficiently low frequencies/temperatures, since the electronic Drude contribution to the layer polarisability diverges at low frequencies, while the optical phonon contribution stays finite. Comparing the two contributions, one can estimate the temperature below which the Drude model is sufficient.

The rest of the paper is organised as follows. In Sec.~\ref{sec:model} we specify the model and sketch the calculation; both are rather standard. In Sec.~\ref{sec:results} we present various regimes of the heat transfer and the associated analytical expressions for the heat current, according to the material properties and experimental conditions.
In Sec.~\ref{sec:experiments} we discuss the relation of our results to the well-studied case of heat transfer between bulk semi-infinite metals, the role of the substrates in the heat transfer, the heat transfer enhancement in the near field, and compare our theory to available experimental results.
All details of calculations are given in three appendices.

\section{The model}
\label{sec:model}


We consider two identical 2D metal sheets held at different temperatures $T_1$ and $T_2$, embedded in vacuum and separated by a gap of width~$d$. A more realistic configuration would be to place a medium with a dielectric constant $\varepsilon$ in the half-space behind each sheet, since in experiments the layers are placed on a substrate. For the sake of simplicity, we focus on $\varepsilon=1$ in most of the paper, and check for the effect of the substrate when specifically needed (see Sec.~\ref{ssec:substrates}). 

We model the metal sheets as infinitely thin layers, characterised by a local 2D Drude conductivity,
\begin{equation}\label{eq:sigmaDrude}
\sigma(\omega) = \frac{\sigma_\text{2D}}{1-i\omega\tau},
\end{equation}
with $\tau$ being the electron momentum relaxation time, assumed to be temperature-independent. This is the case if $\tau$~is determined by elastic scattering on static impurities. Eq.~(\ref{eq:sigmaDrude}) neglects (i)~the spatial dispersion of the conductivity, and (ii)~field variation over the layer thickness. For atomically thin materials, such as doped graphene or transition metal dichalcogenides, condition~(ii) is irrelevant, and condition~(i) holds at distances $d\gg\sqrt{a_\mathrm{2D}\ell}$ ($a_\mathrm{2D}$ and $\ell$ being the 2D screening radius and the electron mean free path, respectively)~\cite{Wise2020}. For thin but macroscopic layers of conventional metals, condition~(ii) imposes that the thickness must be small compared both to the typical wavelength of the waves dominating the heat transfer (which may be rather short for evanescent waves) and to the skin depth at the typical frequency of these waves, while condition~(i) requires the wavelength and the skin depth to be longer than the electron mean free path in the metal.


Our calculation of the heat current between the metals follows the standard FED procedure.
The fluctuating in-plane surface currents $\mathbf{j}^{(\alpha)}(\mathbf{r},t)$ in each sheet obey the fluctuation-dissipation theorem,
\begin{align}
\langle{j}_l(\mathbf{r},t)\,j_m(\mathbf{r}',t')\rangle
=\delta_{lm}&\int\frac{d^2\mathbf{k}\,d\omega}{(2\pi)^3}\,
 \hbar\omega \coth\frac{\hbar\omega}{2 T} \Re\sigma(\omega) \nonumber\\
&{}\times e^{i\mathbf{k}(\mathbf{r}-\mathbf{r}')-i\omega(t-t')},
 \label{eqn:FDT}
\end{align}
where $\textbf{k}$ is the in-plane two-dimensional wavevector and $l,m = x,y$ label the orthogonal in-plane directions and $T=T_1$ or $T_2$. These currents appear as sources in  Maxwell's equations, whose solution in the presence of the conducting sheets determines the fluctuating electric fields $\mathbf{E}(\mathbf{r},t)$. Then, the heat current $J$ (per unit area) from layer~1 to layer~2 is given by the average Joule loss power (per unit area) $\langle\tilde{\mathbf{j}}^{(2)}\cdot\mathbf{E}^{(2)}\rangle-\langle\tilde{\mathbf{j}}^{(1)}\cdot\mathbf{E}^{(1)}\rangle$, where $\tilde{\mathbf{j}}^{(\alpha)}$ is the surface current in layer~$\alpha$, induced by the electric field $\mathbf{E}^{(\alpha)}$ in this layer, which, in turn, is produced by the fluctuating current in the other layer (see Appendix \ref{sec:explicit} for explicit expressions, rather standard).

We emphasize that for thin layers, the Joule losses $\langle\tilde{\mathbf{j}}^{(2)}\cdot\mathbf{E}^{(2)}\rangle-\langle\tilde{\mathbf{j}}^{(1)}\cdot\mathbf{E}^{(1)}\rangle$ are not equal to the average normal component of the Poynting vector in the gap between the layers. The reason is that some part of the radiation emitted by layer 1 may pass through layer 2 and escape to infinity, and vice versa. Whether this escaped radiation should be included in the heat current or not, depends on the precise measurement setup, which may collect this escaped radiation or not. Our calculation thus assumes that the escaped radiation is lost. 
As discussed in the Introduction and in Sec.~\ref{ssec:substrates}, here we focus on the heat transfer from one metal to the other, so we calculate the Joule losses, not the Poynting vector. 
Note that for two semi-infinite metals (the most studied setup), everything is collected inside the metals, so the Poynting vector and the Joule losses match exactly.

In the planar geometry considered here, the solutions of Maxwell's equations are classified by in-plane wave vector $\mathbf{k}$, frequency $\omega$, and two polarisations $p=\mathrm{TE},\mathrm{TM}$ -- transverse electric and transverse magnetic, respectively, for which the electric or the magnetic field vector is parallel to the layers and perpendicular to~$\mathbf{k}$. The contributions to the heat current from modes with different $\mathbf{k},\omega,p$ add up independently, so the heat current $J(T_1,T_2)$ is given by by an integral over $\mathbf{k}$ and $\omega$, and a sum over the polarisations. The integral splits in two contributions: the interior  of the light cone, $\omega>ck$ hosts travelling modes, while in the region $\omega<ck$ the solutions are evanescent.
The resulting heat current is comprised of four additive contributions (TM and TE, travelling and evanescent). Which contribution dominates, depends on the material conductivity, as well as the system temperature and length scales.

In the extreme near field limit, $k\gg\omega/c$, the TM mode field is mostly electric and longitudinal, while the magnetic field is smaller by a factor $\sim\omega/(ck)$; these modes represent the electrostatic coupling by the Coulomb interaction between charge density fluctuations in the two layers. At the same time, for TE modes the field is mostly magnetic, while the electric field is smaller by a factor $\sim\omega/(ck)$; these  modes represent magnetostatic coupling, where the magnetic field established by transverse current fluctuations in one layer drives eddy currents in the second layer. 

In Appendix~\ref{sec:asymptotics} we perform analytically the $\mathbf{k},\omega$ integrals and derive simple asymptotic expressions for the heat current according to the separation and the temperature. For each expression, we can identify the dominant contribution (TM or TE, travelling or evanescent). Our results are approximate; one can describe the heat transfer much more precisely by solving Maxwell's equations for finite-thickness slabs with a material-specific frequency dependence of the conductivity and numerically evaluating the integrals, as routinely done in many works. However, simple approximate expressions (i)~are rather useful when a quick estimate of the heat current is needed, and (ii)~offer a general insight into the dominant physical mechanisms responsible for the heat transfer and enable one to characterise different possibilities.

\section{Results}
\label{sec:results}

\begin{figure*}
\includegraphics[width=0.45\textwidth]{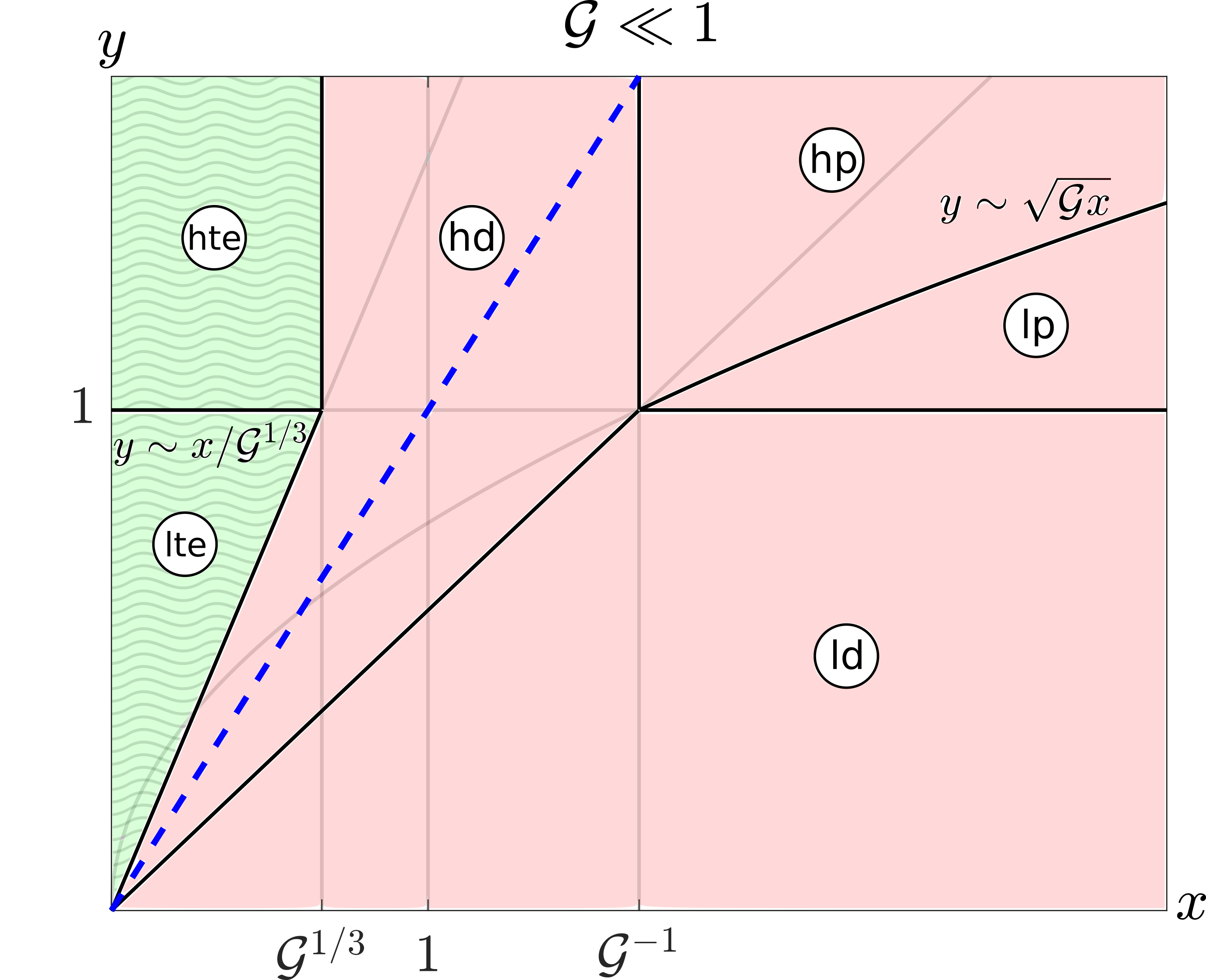}
\hfill
\includegraphics[width=0.45\textwidth]{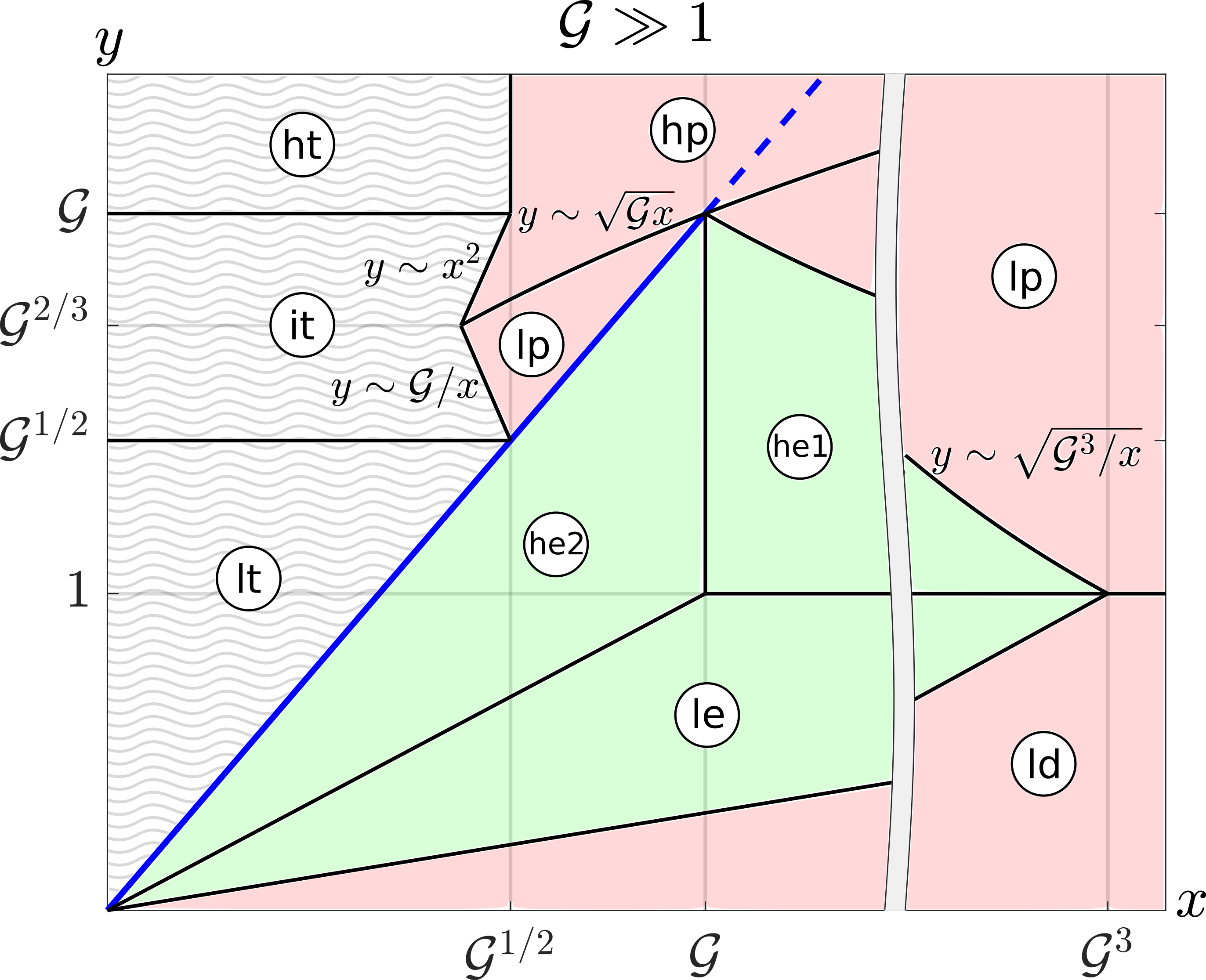}
\caption{
The domains of validity for asymptotic expressions, Eqs.~(\ref{eqn:resultsGsmall}) and (\ref{eqn:resultsGlarge}) in the parameter plane $(1/d,T)$, shown in the dimensionless variables $x\equiv c\tau/d$, $y\equiv T\tau/\hbar$. The crossovers between the regimes are governed by the dimensionless conductivity parameter $\mathcal{G}\equiv 2\pi\sigma_\text{2D}/c$, the left and right panels corresponding to $\mathcal{G}\ll1$ and $\mathcal{G}\gg1$, respectively.
The encircled label of each region corresponds to the subscript at $J(T)$ in Eqs.~(\ref{eqn:resultsGsmall}) and (\ref{eqn:resultsGlarge}). 
Solid lines indicate crossovers between different expressions; straight lines $y/x=\mathrm{const}$ are not labeled for readability (the coefficient can be deduced from the endpoints).
The blue line (solid or dashed) corresponds to $d\sim\lambdabar_T$.
The shading color indicates the heat being predominantly carried by TM modes (red), TE modes (green), or both (white); dominance of travelling waves is indicated by wavy hatching. \\
}
\label{fig:map}
\end{figure*}

For temperature-independent relaxation time, the heat current naturally splits into the difference $J(T_1,T_2) = J(T_1) - J(T_2)$. The detailed analysis of different asymptotic regimes of the $\mathbf{k},\omega$ integrals results in several asymptotic expressions for $J(T)$ in different parametric ranges of~$d$. The magnitudes of the TM and TE travelling contributions are sensitive to the dimensionless conductivity parameter $\mathcal{G} = 2\pi\sigma_\text{2D}/c$ so we proceed to present the results sequentially for small and large values of $\mathcal{G}$.

For $\mathcal{G} \ll 1$, the asymptotic expressions for $J(T)$ are:
\begin{subequations}\label{eqn:resultsGsmall}\begin{eqnarray} 
&& J_{\rm lp}(T)=\frac{\zeta(3)}{4\pi}\,\frac{T^3}{\hbar^2c\mathcal{G}d},
\label{eqn:Jlp}\\ 
&& J_{\rm hp}(T)=\frac{T}{16\pi\tau{d}^2}\,
\mathcal{L}(\mathcal{G}c\tau/d),\label{eqn:Jhp}\\ 
&& J_{\rm ld}(T)=\frac{\zeta(3)}{8\pi}\, \frac{T^3}{\hbar^2c\mathcal{G}d},\label{eqn:Jld}\\  
&& J_{\rm hd}(T)=\frac{1}{16\pi}\,\frac{\mathcal{G}c}{d^3}\,T,\label{eqn:Jhd}\\
&& J_{\rm lte}(T) = \frac{\pi^2}{15} \frac{\mathcal{G}^2}{\hbar^3c^2}\,T^4\ln\frac1{\mathcal{G}}, \label{eqn:Ji} \\
&& J_{\rm hte}(T) = \frac{1}{4\pi}\,\frac{\mathcal{G}^2}{c^2\tau^3}\,T\ln\frac1{\mathcal{G}} ,\label{eqn:Jii}
\end{eqnarray}\end{subequations}
valid in the corresponding regions of the $(1/d,T)$ plane, schematically shown in Fig.~\ref{fig:map}~(left).
The contributions given by Eqs.~(\ref{eqn:Jlp})--(\ref{eqn:Jhd}), with labels corresponding to low-temperature plasmonic, high-temperature plasmonic, low-temperature diffusive, high-temperature diffusive, are the TM evanescent contributions that remain in the Coulomb limit and were calculated in Ref.~\cite{Wise2020}. Equations~(\ref{eqn:Ji}) and (\ref{eqn:Jii}) (with labels corresponding to low-temperature travelling electic and high-temperature travelling electric) are the travelling TE contributions which dominate over the travelling TM contributions by the logarithmic factor $\ln(1/\mathcal{G})$. In Eqs.~(\ref{eqn:resultsGsmall}), $\zeta(x)$ is the Riemann zeta function, and $\mathcal{L}(x)$ is a slow logarithmic function, approximately given by~\cite{Wise2020}
\begin{equation}\label{eqn:Lapprox}
\mathcal{L}(x)\approx\frac{4\ln^3x}{1+(\ln{x})/\ln(1+\ln{x})}.
\end{equation}

For $\mathcal{G} \gg 1$, in addition to the expressions given in 
Eqs.~(\ref{eqn:Jlp})--(\ref{eqn:Jld})
we also have:
\begin{subequations}\label{eqn:resultsGlarge}\begin{eqnarray} 
&& J_{\rm le}(T)=\frac{\pi^2}{15} \frac{\mathcal{G}^2}{\hbar^3 c^2} T^4\ln\frac{\lambdabar_T}{\mathcal{G}d}, \label{eqn:Jle} \\ 
&& J_{\rm he1}(T) = \frac{1}{4\pi} \frac{\mathcal{G}^2}{c^2\tau^3}\,T \ln\frac{c\tau}{\mathcal{G}d}, \label{eqn:Jhe1} \\
&& J_{\rm he2}(T) = \frac{\zeta\left(3\right)}{16\pi} \frac{c}{\mathcal{G} d^3}T, \label{eqn:Jhe2} \\ 
&& J_{\rm lt}(T) = \frac{\pi^2}{45} \frac{T^4}{\hbar^3 c^2 \mathcal{G}}, \label{eqn:Jlt} \\ 
&& J_{\rm it}(T) = \frac{1}{12}\frac{T^2}{\hbar c^2\tau^2}, \label{eqn:Jit} \\ 
&& J_{\rm ht}(T) =  \left(\frac{\sqrt{2}}{12\pi} + \frac{1}{4\pi}\right)  \frac{\mathcal{G}}{c^2\tau^3}T, \label{eqn:Jht} 
\end{eqnarray}\end{subequations}
valid in the corresponding regions of the $(1/d,T)$ plane, schematically shown in Fig.~\ref{fig:map}~(right). 
The contributions given by Eqs.~(\ref{eqn:Jle})--(\ref{eqn:Jhe2}) are the TE evanescent contributions, while Eqs.~(\ref{eqn:Jlt})--(\ref{eqn:Jht}) are the sums of travelling contributions from both polarisations which are of the same order. 


For $\mathcal{G}\gg1$, the travelling channels support resonant Fabry-Perot (FP) modes. In (lt) and (it) regions, many sharp FP modes contribute significantly to the heat current. In the high-temperature case (ht) the FP modes are overdamped since the conductivity $\sigma(\omega)$ becomes small at high frequencies. For temperatures lower than the first mode cutoff energy, $T\ll\pi\hbar{c}/d$, the contributions from the FP modes are exponentially suppressed. However, the prefactor in front of the small thermal exponential turns out to be larger than the evanescent contribution~(\ref{eqn:Jhe2}) in (he2) region. Thus, the FP additive contribution is potentially significant for $c\tau/d<\sqrt{\mathcal{G}}$, where it is dominated by the first FP mode:
\begin{equation}
J_\text{FP1}(T) = 
\frac{\pi{c}T}{[2\mathcal{G}+(\pi{c}\tau/d)^2]d^3}\,e^{-\pi\hbar{c}/(Td)}.
\label{eqn:JFP1}
\end{equation}

In Fig.~\ref{fig:map}, the areas with wavy hatching indicate the regions where the heat transfer is dominated by travelling wave contributions. 
For $\mathcal{G}\ll1$, the evanescent waves dominate at separations up to $d\sim\mathcal{G}^{-1/3}\lambdabar_T$, parametrically larger than the commonly used condition for the near field, $d\ll\lambdabar_T$ [$y\ll x$ in Fig.~\ref{fig:map}~(left)]; the reason for such behaviour is that the low-temperature TM evanescent contribution is determined by $k\ll1/d$ for which the exponential suppression is not efficient.
Moreover, for $\mathcal{G}\ll1$ the near-field transfer is dominated by the TM evanescent contribution, basically, by electrostatic (capacitive) coupling between the two layers. This happens because in a poor conductor, the charge density response is not fast enough to dynamically screen the fluctuating Coulomb field.

For $\mathcal{G}\gg1$, the commonly used inequality $d\ll \lambdabar_T$ does become the accurate condition for evanescent contribution dominance, except for high temperatures where the Drude conductivity is suppressed by high frequency. 
In a large part of the near-field region of the parameter plane the heat current is governed by TE evanescent modes, which correspond to magnetostatic (inductive) coupling between the layers. As discussed in Refs.~\cite{Chapuis2008, Chapuis2008-1} for bulk metals, large conductivity leads to efficient screening of the electric fields, so the magnetostatic coupling becomes more important. The electrostatic coupling takes over only at very short distances or low temperatures, $d\ll\lambdabar_T/(\pi\mathcal{G})^3$, determined by $J_\mathrm{le}(T)\sim{J}_\mathrm{ld}(T)$.

However, for very small~$d$ the finite layer thickness may become important, and/or the assumption of the local response, Eq.~(\ref{eq:sigmaDrude}), may break down. 
Taking, for example, a $10\:\mathrm{nm}$-thick gold film with the bulk plasma frequency $\omega_p = 0.6\times10^{16}\:\mathrm{s}^{-1}$ and relaxation time $\tau = 6\:\mathrm{fs}$~\cite{Ordal1985} gives $\mathcal{G}\approx3.6$.
Since $\lambdabar_T=7.6\:\mu\mbox{m}$ at $T=300~\mbox{K}$, in such structure the crossover to electrostatics occurs at a few nanometers.
We note that in the (ld)~regime, the heat transfer is mainly determined by rather small wave vectors $k\sim(\mathcal{G}\lambdabar_Td)^{-1/2}$ \cite{Wise2020}, so that even at $d=1\:\mbox{nm}$ we obtain $1/k\gtrsim100\:\mbox{nm}$, and the local response assumption should still be formally valid. However, at nanometric distances other physical effects may come into play (electron or phonon tunnelling, surface roughness, etc.), so for ultrathin films of conventional metal we expect the Coulomb mechanism to be relevant mostly at low temperatures.

\section{Discussion}
\label{sec:experiments}

\subsection{Comparison to the bulk case}



The results presented in the previous section show two qualitatively different pictures of the near-field heat transfer between two metallic layers, depending on the value of their dimensionless 2D dc conductivity: for $\mathcal{G}\ll1$, the heat transfer is mostly due to electrostatic coupling between the layers, up to distances significanly exceeding~$\lambdabar_T$, while for $\mathcal{G}\gg1$ the near-field magnetostatic coupling dominates up to distances $d\sim\lambdabar_T$, in close analogy with earlier results on bulk metals. This picture is consistent with the results of Ref.~\cite{Wang2019} where the very same problem of radiative heat transfer between parallel 2D layers was studied numerically. There, a distinction was made between thin and thick metallic films. In this formulation $\mathcal{G}$ is proportional to the layer thickness~$h$ (in the local approximation, the 2D conductivity is simply $\sigma_{2D}=\sigma_{3D}h$, where $\sigma_{3D}$ is the bulk conductivity). In Ref.~\cite{Wang2019}, the heat transfer between two theoretically imagined atomic monolayers of silver, described by a 2D Drude model with $\mathcal{G}\approx2$, is found to be driven by TM evanescent waves, while for thicker films it is TE evanescent waves.

The peculiarity of the 2D geometry is that the 2D conductivity can be compared to two universal scales. One is the speed of light, hence the dimensionless parameter $\mathcal{G}=2\pi\sigma_{2D}/c$ we introduced earlier. The other universal scale is the conductance quantum, $e^2/(2\pi\hbar)$. For $\sigma_{2D}\lesssim{e}^2/(2\pi\hbar)$, or $\mathcal{G}\lesssim{e}^2/(\hbar{c})\approx1/137$, the disorder is too strong, so the metallic conduction is destroyed by localization effects \cite{Altshuler1985, Lee1985}. Thus, the poor conductor regime discussed above, can be realised in the interval $1/137\lesssim\mathcal{G}\ll1$.

The situation is quite different for the bulk metal case. The 3D conductivity $\sigma_{3D}$ has the dimensionality of the inverse time, so that $1/(4\pi\sigma_{3D})$ (in CGS units, while in SI it is $\varepsilon_0/\sigma_{3D}$) has a meaning of the $RC$ time needed to dissolve a charge density perturbation. In conventional metals this time scale is extremely short (in the attosecond range). Still, one can compare $4\pi\sigma_{3D}$ to other scales. One is the electron relaxation time~$\tau$; typically, $4\pi\sigma_{3D}\tau=\omega_p^2\tau^2\gg1$ ($\omega_p$~being the bulk plasma frequency). Moreover, at $T\ll\hbar/\tau$ the relaxation time drops out of the problem, so one cannot construct a dimensionless parameter out of~$\sigma_{3D}$, which could produce different ``asymptotic maps'' of the kind shown in Fig.~\ref{fig:map}. The bulk case turns out to be somewhat similar to the 2D case with $\mathcal{G}\gg1$.

To see the reason for this similarity, let us recall the asymptotic expressions for the heat current between semi-infinite bulk metals, assuming $T\tau\ll\hbar$ (the derivation can be found in Ref.~\cite{Polder1971}, we also give it in Appendix~\ref{app:3D}):
%
%
%
%
\begin{widetext}
\begin{subequations}
\label{eqs:Jbulk}
\begin{align}
J_a(T)={}&{}\frac{\pi^2}{60}\,\frac{\hbar(T/\hbar)^4}{(2\pi\sigma_{3D})^2d^2}
\ln\frac{2\pi\sigma_{3D}}{T/\hbar},
&& d\ll\frac{\delta_T^3}{\lambdabar_T^2}, \label{eq:Ja=}
\\
J_b(T)={}&{}\frac{\zeta(3)}{4\pi^2} \frac{2\pi\sigma_\mathrm{3D} \hbar(T/\hbar)^3}{c^2}, && \frac{\delta_T^3}{\lambdabar_T^2}\ll{d}\ll\delta_T, \label{eq:Jb=}
\\
J_c(T)={}&{}\frac{3\,\zeta(3)}{4\pi^2}\,\frac{c^2T}{2\pi\sigma_{3D}d^4}\ln\frac{d}{\delta_T}, && \delta_T\ll{d}\ll(\lambdabar_T^2\delta_T)^{1/3}, \label{eq:Jc=}
\\
J_d(T)={}&{}\frac{75\,\zeta(7/2) }{256 \sqrt\pi }\,\frac{\hbar(T/\hbar)^{7/2}}{\sqrt{2\pi\sigma_\mathrm{3D}}cd},&& (\lambdabar_T^2\delta_T)^{1/3}\ll{d}\ll\lambdabar_T, \label{eq:Jd=}
\\
J_e(T)={}&{}\frac{35\, \zeta(9/2)}{16\pi^{3/2}} \frac{\hbar(T/\hbar)^{9/2}}{\sqrt{2\pi\sigma_\mathrm{3D}} c^2}, && \lambdabar_T\ll{d}, \label{eq:Je=}
\end{align}
\end{subequations}
\end{widetext}
where the parametric intervals of~$d$ are conveniently defined in terms of two length scales: the thermal wavelength $\lambdabar_T=\hbar{c}/T$ and the normal skin depth at the thermal frequency, $\delta_T={c}/\sqrt{2\pi\sigma_{3D}T/\hbar}\ll\lambdabar_T$. 
The shortest-distance expression~(\ref{eq:Ja=}) is determined by the TM evanescent contribution and corresponds to the Coulomb limit (indeed, it does not contain the speed of light); however, the length scale $\delta_T^3/\lambdabar_T^2$ is extremely short: for $4\pi\sigma_{3D}=10^{17}\:\mbox{s}^{-1}$ at $T=300\:\mbox{K}$, we have $\delta_T=0.22\:\mu\mbox{m}$ and $\lambdabar_T=7.6\:\mu\mbox{m}$, so $\delta_T^3/\lambdabar_T^2\sim2\:\mbox{\AA}$ and becomes even smaller at lower temperatures, invalidating the local approximation and making Eq.~(\ref{eq:Ja=}) irrelevant for conventional metals.
Equations~(\ref{eq:Jb=}) and~(\ref{eq:Jc=}) originate from the TE evanescent contribution and correspond to magnetostatic coupling~\cite{Chapuis2008}.
Equation~(\ref{eq:Jd=}) contains both TM evanescent and TM travelling contributions which are of the same order at such distances (only the evanescent one was evaluated in Ref.~\cite{Polder1971}); in fact, for both contributions the integral is dominated by wave vectors~$k$ very close to $\omega/c$, and the fields vary weakly across the gap so there is no sharp physical distinction between travelling and evanescent waves. Finally, Eq.~(\ref{eq:Je=}) comes from the TE and TM travelling waves and is contributed by many Fabry-Perot modes inside the gap.

It is easy to see that by the order of magnitude, Eqs.~(\ref{eq:Jb=}), (\ref{eq:Jc=}), (\ref{eq:Jd=}) and (\ref{eq:Je=}) can be obtained from Eqs.~(\ref{eqn:Jle}), (\ref{eqn:Jhe2}), (\ref{eqn:Jld}) and (\ref{eqn:Jlt}), respectively, by  replacing $\mathcal{G}=2\pi\sigma_{2D}/c\to2\pi\sigma_{3D}\delta_\omega/c$, where the skin depth $\delta_\omega=c/\sqrt{2\pi\sigma_{3D}\omega}$ corresponds to the typical frequency scale determining the integral: it is $\delta_T$ for Eqs.~(\ref{eq:Jb=}), (\ref{eq:Jd=}) and (\ref{eq:Je=}), determined by frequencies $\omega\sim{T}/\hbar$, and $\delta_\omega\sim{d}$ for Eq.~(\ref{eq:Jc=}), where the frequency integral is logarithmic, with the lower cutoff corresponding to $\delta_\omega\sim{d}$ (see Appendix~\ref{app:3D} for details). 
This replacement roughly corresponds to modelling the semi-infinite metal as an effective metallic layer whose thickness corresponds to the field penetration depth. Such effective layer is characterised by the dimensionless $\mathcal{G}_\textrm{eff}\sim\sqrt{2\pi\sigma_{3D}/\omega}$, so that $\mathcal{G}_\textrm{eff}\gg1$ for conventional metals and reasonable temperatures.
We note that this effective layer analogy should be used with caution, since the frequency dependence of $\delta_\omega$ sometimes makes the convergence scale of the frequency integral different from the case of fixed layer thickness.

\subsection{Role of the substrates}
\label{ssec:substrates}

The expressions given in Sec.~\ref{sec:results} correspond to the heat transferred from one metallic sheet to the other, not including the radiation transmitted behind each sheet. In an experiment, this transmitted radiation can be absorbed by dielectric substrates  (even if the absorption by the dielectric material is very weak, the transmitted radiation can still be absorbed if the substrate is thick enough) or captured by some background parts of the structure. Whether the transmitted radiation should be included in the measured heat current or not, depends on the specific measurement scheme. The measurement can be done directly on the metallic layers, as, e.~g., in Ref.~\cite{Kralik2012}; the radiation absorbed in the substrate leads to a very weak heating of the latter since this absorption occurs in a large volume, and has little effect on the metallic layers. The opposite example is Ref.~\cite{Yang2018}, where the measurement was actually performed behind the substrate, so that all radiation was collected, and good thermal contact between graphene sheets and the substrate was ensured.

If radiation absorbed by the thick dielectric substrate is included, one should also include radiation emitted by the substrate, which is equivalent to adding an incident black-body heat flux
\begin{equation}\label{eqn:PB=}
J_\mathrm{bb}(T) = \frac{\pi^2}{60}\,\frac{T^4}{c^2\hbar^3}.
\end{equation}
Its effect is especially important for $\mathcal{G}\ll1$ since the transmission of each sheet is close to unity in this case. For thick dielectric substrates with dielectric constant $\varepsilon=1$ and an infinitesimal imaginary part, almost all incident black-body heat flux is transmitted through the sheets and absorbed on the other side, so the far-field expressions $J_\mathrm{lte}(T)$ and $J_\mathrm{hte}(T)$, Eqs.~(\ref{eqn:Ji}) and (\ref{eqn:Jii}), should be replaced by Eq.~(\ref{eqn:PB=}). This starts to dominate over the near-field contribution $J_\mathrm{hd}(T)$ at shorter distances, $d\sim\lambdabar_T\mathcal{G}^{1/3}\ll\lambdabar_T$. This is natural, since the near field contribution is still determined by the sheets, while the far field transfer is essentially between the substrates.

For $\mathcal{G}\gg1$, the low-temperature far-field expression~(\ref{eqn:Jlt}) remains valid, since the layer transmission is too small. At intermediate temperatures,  $\sqrt{\mathcal{G}}\ll{T}\tau/\hbar\ll\mathcal{G}$, the sheet transmission is still small, but it is already larger than the absorption, so the far field contribution is determined by the fraction of the black-body radiation entering the Fabry-Perot resonator, $J(T)\sim{J}_\mathrm{bb}(T)\,(T\tau/\hbar{G})^2$, which is larger than $J_\mathrm{it}(T)$, Eq.~(\ref{eqn:Jit}). At $T\tau/\hbar\gg1$ the conductivity at relevant frequencies is so small, that the transmission of the layers is close to~1, and instead of Eq.~(\ref{eqn:Jht}) the far-field heat current is the black-body one, Eq.~(\ref{eqn:PB=}).

\subsection{Near-field enhancement of the heat transfer}

When studying radiative heat transfer between objects, one is often interested in comparing it to the radiative transfer between black bodies of the same geometry. In the planar geometry considered here, the black-body heat current is given by Eq.~(\ref{eqn:PB=}) and does not depend on~$d$. Metals are not perfect emitters/absorbers, so in the far field they exchange less heat than black bodies. This is seen by comparing the far-field expressions (\ref{eqn:Ji}), (\ref{eqn:Jii}), (\ref{eqn:Jlt})--(\ref{eqn:Jht}), which are all $d$~independent,  to Eq.~(\ref{eqn:PB=}). At the lowest temperatures, we have $J_\mathrm{lte}(T)/J_\mathrm{bb}(T)=4\mathcal{G}^2\ln(1/\mathcal{G})$ and $J_\mathrm{lt}(T)/J_\mathrm{bb}(T)=4/(3\mathcal{G})$, for $\mathcal{G}\ll1$ and $\mathcal{G}\gg1$, respectively. At higher temperatures, even smaller values are obtained. Only at $\mathcal{G}\sim1$ the metallic sheets approach the black-body limit in the far field, due to impedance matching with vacuum.

However, it is well known that the coupling of evanescent modes can lead to significant, $d$-dependent contributions to the heat transfer between closely spaced conducting bodies, resulting in an overall enhancement of the radiative power compared to the black-body result (\cite{Joulain2005, Volokitin2007, Song2015, Biehs2020} and references therein). So there are two competing effects: the far field contribution is weaker than that of black bodies due to metals being imperfect emitters, meanwhile between metals there is an extra contribution from the evanescent waves that dominates in the near field (evanescent waves do not contribute to black-body radiation into the vacuum). To assess when the near field contribution leads to an enhancement over the black-body result, one needs to compare various near-field expressions in Sec.~\ref{sec:results} to Eq.~(\ref{eqn:PB=}). For example, at the lowest temperatures, $J_\mathrm{hd}(T)$ and $J_\mathrm{he2}(T)$ [Eqs.~(\ref{eqn:Jhd}) and~(\ref{eqn:Jhe2})] overcome the black-body current at $d\lesssim\lambdabar_T\mathcal{G}^{1/3}$ and  $d\lesssim\lambdabar_T\mathcal{G}^{-1/3}$, respectively, for  $\mathcal{G}\ll1$ and $\mathcal{G}\gg1$.

The strongest enhacement is obtained at small separations (since the near-field contribution always grows with decreasing~$d$) and low temperatures (since the black-body expression has the highest power of temperature). Thus, we need the ratio of Eqs.~(\ref{eqn:Jld}) and~(\ref{eqn:PB=}):
\begin{equation}
\frac{J_\mathrm{ld}(T)}{J_\mathrm{bb}(T)}=\frac{15\,\zeta(3)}{2\pi^3}\,\frac{1}{\mathcal{G}}\,
\frac{\lambdabar_T}{d}.
\end{equation}
Note that the enhancement is stronger for smaller~$\mathcal{G}$; indeed, in this regime the near-field transfer is dominated by the Coulomb interaction which is screened less efficiently in poorly conducting metals. Taking $T=300\:\mbox{K}$, $d=10\:\mbox{nm}$, and $\mathcal{G}=0.01$ (we remind that for smaller values of $\mathcal{G}$ the Drude description is not valid), we obtain the ratio of $2\times10^4$.
For bulk metals, the relevant ratio is $J_b(T)/J_\mathrm{bb}(T)\approx0.19\,(4\pi\hbar\sigma_{3D}/T)$ [Eq.~(\ref{eq:Jb=}), since Eq.~(\ref{eq:Ja=}) becomes valid at unrealistically short distances], which amounts to about $3\times10^4$ for $4\pi\sigma_{3D}=7\times10^{18}\:\mbox{s}^{-1}$ (silver at room temperature).

\subsection{Comparison to experiments}


Values $\mathcal{G}\lesssim1$ are characteristic of atomically thin 2D materials. This is illustrated by a recent experiment~\cite{Yang2018}, where two doped monolayer graphene sheets were placed on insulating silicon ($\varepsilon=11.7$) and separated by a 400~nm wide vacuum gap. The Fermi energy of $0.27\:\mbox{eV}$ and the relaxation time $\tau=100\:\mbox{fs}$ give $\mathcal{G}=0.6$. The linear thermal conductance per unit area $dJ/dT=30\:\mbox{W}\,\mbox{m}^{-2}\,\mbox{K}^{-1}$ was measured around room temperature. These conditions correspond to the high-temperature plasmon regime, Eq.~(\ref{eqn:Jhp}), where the substrate dielectric constant $\varepsilon$ enters only inside the logarithmic function~$\mathcal{L}$~\cite{Wise2020}. Setting $\mathcal{L}=1$ in Eq.~(\ref{eqn:Jhp}) gives $dJ/dT=11\:\mbox{W}\,\mbox{m}^{-2}\,\mbox{K}^{-1}$, which agrees by order of magnitude with the experimental value.

Thin layers of conventional metals are typically characterised by $\mathcal{G}\gg1$. Several experiments have been reported in the literature. In each case, it is important to compare the layer thickness~$h$ to the the skin depth $\delta_\omega$ at the relevant frequency, to ensure the layers should correspond to the 2D limit, rather than the bulk one (the latter being the case of Refs.~\cite{Hargreaves1969,Song2016}).

Heat transfer in a wide range of interlayer separations and temperatures was studied in Ref.~\cite{Kralik2012} for two $150 \,\mathrm{nm}$ thick tungsten layers on alumina substrates. The measured dc conductivity of the material $4\pi\sigma_\mathrm{3D} = 0.6\times 10^{18}\,\mathrm{s}^{-1}$ (constant in the temperature range of the experiment) corresponds to a value of the dimensionless conductivity parameter $\mathcal{G} \approx 150$. The skin depth at $T=40\:\mbox{K}$ is $\delta_T = 240\:\mbox{nm}$, and even longer at lower temperatures, so the layers are close to the 2D limit.
The separation between the layers was varied over $d=1-300\,\mathrm{\mu m}$, while the temperatures were $T_1 = 5\,\mathrm{K}$ and $T_2 = 10-40\,\mathrm{K}$, corresponding to regions (he2) and (lt) in Fig.~\ref{fig:map}~(right). It can be easily checked that in these regions, the dielectric substrate plays no role as long as $\sqrt{\varepsilon} \ll \mathcal{G}$, which clearly holds here. Although the numerical calculation accounting for the finite layer thickness does better in closely matching the experimental points (see Fig.~2 of Ref.~\cite{Kralik2012}), our simple expressions~(\ref{eqn:Jhe2}) and~(\ref{eqn:Jlt}) (i)~agree with the observed values within a factor of~3 without any fitting parameters, (ii)~give the correct distance dependence throughout the experiment, (iii)~capture the observed approximate collapse of the rescaled data for $J(T)/T^4$ on a function of a single variable $Td$, and (iv)~correctly predict the separation $d\approx0.5\,\lambdabar_T$, at which the crossover between the near-field and the far-field regimes occurs, $J_\mathrm{he2}(T)=J_\mathrm{le}(T)$.

A recent publication \cite{Sabbaghi2020} presents measurements of heat transfer between two aluminium films of varying thicknesses $h = 13-79\,\mathrm{nm}$, separated by a fixed vacuum gap $d=215\,\mathrm{nm}$ and attached to silicon substrates. The experiment was performed around room temperature with one film being heated such that $\Delta T = 25-65\,\mathrm{K}$. 
Taking the values $\omega_p = 1.93\times10^{16}\,\mathrm{s^{-1}}$ and $\tau \approx 5\,\mathrm{fs}$ \cite{Modest2013} used in Ref.~\cite{Sabbaghi2020} to interpret the data, we obtain 
$\delta_T\approx50\:\mbox{nm}$ and $\mathcal{G}\approx40$ for the thinnest layer with $h=13\:\mbox{nm}$.
Then Eq.~(\ref{eqn:Jhe2}) predicts $dJ/dT = 250 \,\mathrm{W\, m^{-2}/K}$, which agrees in order of magnitude with the reported value, $dJ/dT = 60 \,\mathrm{W\, m^{-2}/K}$.

An intriguing feature of the results reported in Ref.~\cite{Sabbaghi2020} is the independence of $dJ/dT$ of the layer thickness. This agrees neither with our 2D expressions, nor with the more precise simulations done in Ref.~\cite{Sabbaghi2020}. All theoretical results point to a non-monotonic dependence of the heat current on the layer thickness or dc conductivity [the latter is also true for the bulk limit expressions~(\ref{eqs:Jbulk})]. Further experimental investigations of this dependence would be interesting.

\section{Conclusions}

In this paper, we have performed an analytical calculation of the radiative heat current between two thin metallic layers, using the standard framework of fluctuational electrodynamics and a local 2D Drude model for the electromagnetic response of each layer.
We have identified two different classes of such structures, distinguished by the dimensionless 2D dc conductivity $\mathcal{G}=2\pi\sigma_{2D}/c$. For poor conductors with $\mathcal{G}\ll 1$, typically represented by atomically thin 2D materials, the heat transfer is dominated by evanescent modes at distances~$d$ extending well beyond~$\lambdabar_T$, and the main coupling mechanism in this near-field regime is the Coulomb interaction between electrons in the two layers. Good conductors with $\mathcal{G}\gg1$, such as thin films of conventional metals, behave more similarly to the bulk limit, studied in earlier works: the crossover from near to far field occurs at $d\sim\lambdabar_T$ at not too high temperatures, and the near-field transfer is dominated by magnetostatic (inductive) coupling between the layers in a wide range of parameters.

We have derived several simple approximate asymptotic expressions for the heat current valid in different parametric ranges of interlayer separation distance and temperature. Comparing these expressions with the available experimental data, we saw that they give valid order-of-magnitude estimates of the heat current and correctly capture its dependence on the distance and temperature. Better agrreement with the experimental results can be reached by a more detailed modelling of each system geometry and the dielectric response, which is strongly system-specific and lies beyond the scope of our work.
Still, our approximate results offer a useful insight into the main physical mechanisms responsible for the heat transfer. 

\acknowledgements
We thank J.-J. Greffet, J. Pekola, B.~Van~Tiggelen, and C.~Winkelmann for helpful and stimulating discussions.
This project received funding from the European Union's Horizon 2020 research and innovation programme under the Marie Sk{\l}odowska-Curie Grant Agreement No. 766025.


\appendix

\section{Explicit general expression for the heat current between two thin metallic sheets}
\label{sec:explicit}

We solve Maxwell's equations for the monochromatic components of the electric and magnetic field, $\mathbf{E}_{\mathbf{k}\omega}(z)\,e^{i\mathbf{k}\mathbf{r}-i\omega{t}}$ and $\mathbf{B}_{\mathbf{k}\omega}(z)\,e^{i\mathbf{k}\mathbf{r}-i\omega{t}}$ in the planar geometry with the two metallic sheets placed at $z=z_1,z_2$ with $z_2-z_1=d$, while the position-dependent dielectric constant $\varepsilon(z)$ accounts for whatever (non-magnetic, isotropic) dielectric medium surrounds the layers:
\begin{align}
\left(i\mathbf{k}+\mathbf{e}_z\,\frac\partial{\partial{z}}\right)
\times\mathbf{E}_{\mathbf{k}\omega}={}&{}\frac{i\omega}{c}\,\mathbf{B}_{\mathbf{k}\omega},\\
\left(i\mathbf{k}+\mathbf{e}_z\,\frac\partial{\partial{z}}\right)
\times\mathbf{B}_{\mathbf{k}\omega}={}&\frac{4\pi}c\sum_{\alpha=1,2}\delta(z-z_\alpha)
\left(\mathbf{j}_{\mathbf{k}\omega}^{(\alpha)}+\tilde{\mathbf{j}}_{\mathbf{k}\omega}^{(\alpha)}\right)\nonumber\\
{}&{}-\frac{i\omega}{c}\,\varepsilon(z)\,\mathbf{E}_{\mathbf{k}\omega},
\end{align}
where $\mathbf{e}_z$ is the unit vector in the $z$ direction, perpendicular to the layers.
The surface current in each layer $\alpha=1,2$ consists of two contributions: $\tilde{\mathbf{j}}_{\mathbf{k}\omega}^{(\alpha)}=\sigma_\alpha(\omega)\,\mathbf{E}_{\mathbf{k}\omega}(z_\alpha)$ is the induced current due to the electric field, while the fluctuating currents $\mathbf{j}_{\mathbf{k}\omega}^{(\alpha)}=(\mathbf{j}_{-\mathbf{k},-\omega}^{(\alpha)})^*$ are complex Gaussian random variables with the correlator determined by the fluctuation-dissipation theorem~(\ref{eqn:FDT}):
\begin{align}
\langle{j}_{\mathbf{k}\omega,l}^{(\alpha)}\,{j}_{\mathbf{k}'\omega',m}^{(\alpha')}\rangle
={}&{}(2\pi)^3\,\delta(\mathbf{k}+\mathbf{k}')\,\delta(\omega+\omega')\,\delta_{\alpha\alpha'}\delta_{lm}\nonumber\\
{}&{}\times
\hbar\omega \coth\frac{\hbar\omega}{2 T_\alpha} \Re\sigma_\alpha(\omega)
\end{align}
Because of $\delta_{lm}$ on the right-hand side of this equation, current fluctuations are independent for any two orthogonal directions, so it is convenient to pass to the longitudinal and transverse basis ($p$ and $s$ polarisations, respectively):
\begin{equation}
\mathbf{j}_{\mathbf{k}\omega}^{(\alpha)}=j_{\mathbf{k}\omega}^{(\alpha)}\,\frac{\mathbf{k}}{k}
+j_{\mathbf{k}\omega}^{(\alpha)}\,\frac{\mathbf{e}_z\times\mathbf{k}}{k}.
\end{equation}
In this basis the solutions of Maxwell's equations decouple into transverse magnetic (TM) and transverse electric (TE) modes, whose contribution to the heat current is simply additive.

To model different metal sheets mounted on identical dielectric substrates separated by vacuum, we take $\varepsilon(z_1<z<z_2)=1$, $\varepsilon(z<z_1)=\varepsilon(z>z_2)=\varepsilon>1$.
This leads to the spatial dependence of the electric and magnetic fields $\propto e^{i\mathbf{k}\mathbf{r}\pm{i}q_zz}$ for $z_1<z<z_2$, and $\propto e^{i\mathbf{k}\mathbf{r}+iq_z'z},e^{i\mathbf{k}\mathbf{r}-iq_z'z}$ for $z>z_2$ and $z<z_1$, respectively. Here we defined 
\begin{subequations}\begin{eqnarray}
&&q_z=\left\{\begin{array}{ll}
\sqrt{\omega^2/c^2-k^2}\,\sign\omega,&|\omega|>ck,\\
i\sqrt{k^2-\omega^2/c^2},&|\omega|<ck,
\end{array}\right.
\\ &&
q_z'=\left\{\begin{array}{ll}
\sqrt{\varepsilon\omega^2/c^2-k^2}\,\sign\omega,&\sqrt{\varepsilon}|\omega|>ck,\\
i\sqrt{k^2-\varepsilon\omega^2/c^2},&\sqrt{\varepsilon}|\omega|<ck.
\end{array}\right.
\end{eqnarray}\end{subequations}
At $|\omega|>ck$, the metallic layers are coupled by travelling waves, while for $|\omega|<ck$ the solutions in the gap are evanescent waves, where the fields' strength decays away from the layers. The solutions are matched at $z=z_1$ and $z=z_2$ using the standard boundary conditions: continuity of the in-plane component of the electric field $\mathbf{E}_\|$, and a jump in the magnetic field in-plane component, determined by the total surface current (the fluctuatinng sources as well as the induced current $\sigma\mathbf{E}_\|$).

The heat current from, say, sheet 1 to the sheet 2 is given by the average Joule loss power per unit area, $J(T_1, T_2) = \langle\tilde{\mathbf{j}}^{(2)}\cdot\mathbf{E}_\|(z_2)\rangle-\langle\tilde{\mathbf{j}}^{(1)}\cdot\mathbf{E}_\|(z_1)\rangle$, determined unambiguously due to the continuity of $\mathbf{E}_\|(z)$. For a temperature independent relaxation time this heat current splits into $J(T_1, T_2) = J(T_1)-J(T_2)$, where
\begin{equation}\label{eqn:J(T)=}
J(T)=\int\limits_0^\infty\frac{d\omega}{2\pi}\frac{\hbar\omega}{e^{\hbar\omega/T}-1}
\int\!\frac{d^2\mathbf{k}}{(2\pi)^2}
\sum_{j=p,s}\frac{a_{1j}a_{2j}\,|e^{iq_zd}|^2}{|1-r_{1j}r_{2j}e^{2iq_zd}|^2}
\end{equation}
is expressed in terms of reflectivities $r_{\alpha{j}}$ and emissivities $a_{\alpha{j}}$ for the $p$ and $s$ polarisations:
\begin{subequations}\begin{eqnarray}
&&r_{\alpha{p}}=\frac{q_z-q_z'/\varepsilon
+4\pi\sigma_\alpha q_zq_z'/(\varepsilon\omega)}{q_z+q_z'/\varepsilon+4\pi\sigma_\alpha q_zq_z'/(\varepsilon\omega)},\\
&&a_{\alpha{p}}=\frac{4|q_z||q_z'/\varepsilon|^2(4\pi\Re\sigma_\alpha/\omega)}%
{|q_z+q_z'/\varepsilon+4\pi\sigma_\alpha q_zq_z'/(\varepsilon\omega)|^2},\label{eqn:emissp}\\
&&r_{\alpha{s}}=\frac{q_z-q_z'
-4\pi\omega\sigma_\alpha/c^2}{q_z+q_z'+4\pi\omega\sigma_\alpha/c^2},\\
&&a_{\alpha{s}}=\frac{4|q_z|(4\pi\omega\Re\sigma_\alpha/c^2)}%
{|q_z+q_z'+4\pi\omega\sigma_\alpha/c^2|^2}\label{eqn:emisss}.
\end{eqnarray}\end{subequations}
The emissivities can also be written as
\begin{subequations}\label{eqn:emissreflect}\begin{align}
a_{\alpha{p}}={}&{}(1-|r_{\alpha{p}}|^2)\,\theta(|\omega|-ck)
+2\Im{r}_{\alpha{p}}\,\theta(ck-|\omega|)\nonumber\\
{}&{}-\left|\frac{q_z'}{\varepsilon{q}_z}\right|\,|t_{\alpha{p}}|^2\,
\theta(\sqrt{\varepsilon}|\omega|-ck),\\
a_{\alpha{s}}={}&{}(1-|r_{\alpha{s}}|^2)\,\theta(|\omega|-ck)
+2\Im{r}_{\alpha{s}}\,\theta(ck-|\omega|)\nonumber\\
{}&{}-\left|\frac{q_z'}{{q}_z}\right|\,|t_{\alpha{s}}|^2\,
\theta(\sqrt{\varepsilon}|\omega|-ck),
\end{align}\end{subequations}
where $\theta(x)$ is the Heaviside step function, and $t_{\alpha{j}}$ are the transmittivities:
\begin{subequations}\begin{eqnarray}
&&t_{\alpha{p}}=\frac{2q_z}{q_z+q_z'/\varepsilon+4\pi\sigma_\alpha q_zq_z'/(\varepsilon\omega)},\\
&&t_{\alpha{s}}=\frac{2q_z}{q_z+q_z'+4\pi\omega\sigma_\alpha/c^2}.
\end{eqnarray}\end{subequations}
Note the difference between Eqs.~(\ref{eqn:emissreflect}) and Eq.~(2) of Ref.~\cite{Wang2019}, where the third term is absent in both polarisations. Without the third term, Eq.~(\ref{eqn:J(T)=}) gives the average value of the Poynting vector in the gap between the two layers, and also counts the heat flux which is not absorbed by the metal, but irradiated to infinity behind it, due to the finite transmission. Eqs.~(\ref{eqn:emissreflect}) without the third term originally appeared in Ref.~\cite{Volokitin2001} for the problem of heat transfer between two semi-infinite materials. In that geometry, all heat flux transmitted through the surface is eventually absorbed by the material. In the thin layer geometry, whether the transmitted flux is detected or not, depends on the specific experimental measurement scheme. In our calculation, we assume that the transmitted radiation is lost, and thus use the full Eqs.~(\ref{eqn:emissreflect}).


\section{Derivation of asymptotic expressions for the heat current between two thin metallic sheets}
\label{sec:asymptotics}

Here we derive asymptotic expressions for $J(T)$ in the specific case of identical sheets embedded in vacuum [$\sigma_1(\omega) = \sigma_2(\omega)$ and $\varepsilon = 1$] and compute separately the travelling and evanescent wave contributions for each of the two polarisations. We quantify the contribution made by each wave type and polarisation in each region of the $(1/d,T)$ parameter plane, before comparing the size of the additive contributions and identifying which are dominant. It is convenient to introduce the dimensionless parameters $x\equiv{c}\tau/d$ and $y\equiv{T}\tau/\hbar$, as well as dimensionless integration variables: $\xi=|q_z|c\tau$ instead of $k$ [noting that $k\,dk=\xi\,d\xi/\left(c\tau\right)^2$], and $\eta=\omega\tau$. For the travelling waves, the integration is over the region $0<\xi<\eta<\infty$, while for the evanescent waves it is $0<\xi,\eta<\infty$.

\subsection{TM travelling contribution}

In the dimensionless variables, the TM travelling contribution to Eq.~(\ref{eqn:J(T)=}) can be rewritten exactly as
\begin{subequations}\begin{align}
&J_\mathrm{TM}^\mathrm{t}=\frac{\hbar\mathcal{G}^2}{\pi^2c^2\tau^4}
\int_0^\infty\frac{\eta^3\,d\eta}{e^{\eta/y}-1}\,
\int_0^\eta\frac{\xi^3\,d\xi}{D^p_+D^p_-},\label{eq:TMradiative}\\
&D^p_\pm\equiv|\eta(1-i\eta)+\mathcal{G}\xi(1\pm{e}^{i\xi/x})|^2.
\label{eqn:DpmTM=}
\end{align}\end{subequations}
The case $\mathcal{G}\ll1$ is very simple to handle, since for $\varepsilon=1$ one can neglect the reflection coefficients in the denominator of Eq.~(\ref{eqn:J(T)=}), and simply set $\mathcal{G}\to0$ in Eq.~(\ref{eqn:DpmTM=}), since $\xi<\eta$. This gives
\begin{align}
J_\mathrm{TM}^\mathrm{t}={}&{}\frac{\hbar\sigma_\mathrm{2D}^2}{c^4\tau^4}
\int_0^\infty
\frac{\eta^3\,d\eta}{(1+\eta^2)^2(e^{\eta/y}-1)}\nonumber\\
={}&{}\left\{\begin{array}{ll}
\pi^2\mathcal{G}^2T^4/(60\hbar^3c^2), & y\ll1,\\
\mathcal{G}^2T/(16\pi{c}^2\tau^3), & y\gg1. \label{eq:TMtsmallG=}
\end{array}\right.
\end{align}
For $\mathcal{G}\gg1$, each layer behaves at low frequency as a well-reflecting mirror, so the structure may host Fabry-Perot modes.
The Fabry-Perot modes manifest themselves as deep minima in $D^p_\pm$ at specific values of $\xi/x=\pi,2\pi,3\pi,\ldots$. These minima are important when $\mathcal{G}\xi\gg\eta\sqrt{1+\eta^2}$, which is precisely the condition of good reflection. Thus, a much more elaborate analysis is needed to evaluate the integral. 

Let us focus on the contributions from the region $\xi\gg{x}$, when many modes contribute, and even if they are overdamped, $e^{i\xi/x}$ oscillates fast. In the general case~(\ref{eqn:J(T)=}) we average over the fast oscillations in the denominator which leads to the simple replacement \cite{Fu2006}:
\begin{equation}
\frac{a_{1j}a_{2j}}{\left|1-r_{1j}r_{2j}e^{2iq_zd}\right|^2} \rightarrow \frac{a_{1j}a_{2j}}{1-\left|r_{1j}\right|^2\left|r_{2j}\right|^2}, \label{eq:fastoscillation}
\end{equation}
valid as long as $a_{\alpha j}$ and $r_{\alpha j}$ are smooth functions of $q_z$ on the scale $q_z\sim1/d$.


Applying this averaging to the contribution in Eq.~(\ref{eq:TMradiative}) leads to 
\begin{align}
J_\mathrm{TM}^\mathrm{t}={}&{}\frac{\hbar\mathcal{G}^2}{\pi^2c^2\tau^4}
\int_0^\infty\frac{\eta^3\,d\eta}{e^{\eta/y}-1}\nonumber\\
{}&{}\times\int\limits_0^\eta\frac{\xi^3\,d\xi}%
{[\eta^2+\eta^4+2\mathcal{G}\xi\eta+2(\mathcal{G}\xi)^2][\eta^2+\eta^4+2\mathcal{G}\xi\eta]}.
\end{align}
Note that $x$ dropped out, and enters only through the condition $\xi\gg{x}$. Note also that the $\xi$ integral is always determined by the upper limit $\xi\sim\eta$. As for the $\eta$ integral, it may converge at $\eta\sim{y}$ when cut off by the Bose function, or, for too large~$y$, it may be cut off by other factors in the denominator at some $\eta\ll{y}$. In this latter case, one can expand the exponential in the Bose function, which becomes just $y/\eta$. We can identify three regions in~$y$.\\
(i) For $y\ll\sqrt{\mathcal{G}}$, the integrals separate and converge at $\xi\sim\eta\sim{y}$, so the fast oscillation condition is $x\ll{y}$:
\begin{equation}
J_\mathrm{TM}^\mathrm{t}=\frac{\hbar\mathcal{G}^2}{\pi^2c^2\tau^4}
\int_0^\infty\frac{\eta^3\,d\eta}{e^{\eta/y}-1}\,
\int_0^\eta\frac{\xi^3\,d\xi}{4(\mathcal{G}\xi)^3\eta}
=\frac{\pi^2T^4}{60\,\hbar^3c^2\mathcal{G}}.\label{eq:FP1=}
\end{equation}
(ii) For $\sqrt{\mathcal{G}}\ll{y}\ll\mathcal{G}$, we keep $2(\mathcal{G}\xi)^2$ in the first bracket and $\eta^4$ in the second one (again, oscillations are fast when $x\ll{y}$):
\begin{equation}
J_\mathrm{TM}^\mathrm{t}=\frac{\hbar\mathcal{G}^2}{\pi^2c^2\tau^4}
\int_0^\infty\frac{\eta^3\,d\eta}{e^{\eta/y}-1}\,
\int_0^\eta\frac{\xi^3\,d\xi}{2(\mathcal{G}\xi)^2\eta^4}
=\frac{T^2}{24\,\hbar{c}^2\tau^2}.\label{eq:FP2=}
\end{equation}
(iii) For $y\gg\mathcal{G}$, we expand the Bose function, the integral converges at $\eta\sim\mathcal{G}$ (it is convenient to write $\xi=u\eta$);  the oscillations are fast when $x\ll\mathcal{G}$:
\begin{align}
J_\mathrm{TM}^\mathrm{t}={}&{}\frac{\hbar\mathcal{G}^2}{\pi^2c^2\tau^4}
\int_0^\infty{y}\eta^2\,d\eta
\int_0^1\frac{\eta^4u^3\,du}{[\eta^4+2(\mathcal{G}\eta)^2u^2]\eta^4}\nonumber\\
={}&{}\frac{\sqrt{2}\,\mathcal{G}T}{12\pi{c}^2\tau^3}.
\end{align}

Let us now pick the contributions from $\xi\ll{x}$. Then, $e^{i\xi/x}$ can be expanded (we again  write $\xi=u\eta$):
\begin{align}
J_\mathrm{TM}^\mathrm{t}={}&{}\frac{\hbar\mathcal{G}^2}{\pi^2c^2\tau^4}\int_0^\infty
\frac{\eta^3\,d\eta}{e^{\eta/y}-1}\times\nonumber\\
{}&{}\times\int_0^1\frac{u^3\,du}%
{[(1+2\mathcal{G}u)^2+\eta^2][1+\eta^2(1+\mathcal{G}u^2/x)^2]}.
\end{align}
There are three possible cutoff scales for $\eta$: $y$, $1+2\mathcal{G}u$, and $(1+\mathcal{G}u^2/x)^{-1}$. Which one of the three is effective, depends on the positioning of~$y$ with repect to other scales. Again, three cases arise.\\
(iv) For $y\ll1$, we can neglect $\eta^2$ in the first bracket in the denominator, so the $\eta$ integral converges at $\eta\sim{y}$. In the second  bracket, $\eta^2$ plays a role only if $\mathcal{G}{u}^2/x\gg1$, so the second bracket can be approximated as $1+(\mathcal{G}u^2\eta/x)^2$ for any $\mathcal{G}{u}^2/x$. We also assume that $\mathcal{G}u\gg1$, which will be verified afterwards. Then the denominator becomes $4\mathcal{G}^2{u}^2[1+(\mathcal{G}{u}^2\eta/x)^2]$, so the ${u}$ integral converges at ${u}\sim\min\{1,\sqrt{x/(\mathcal{G}y)}\}$, giving
\begin{align}
J_\mathrm{TM}^\mathrm{t}={}&{}\frac\hbar{8\pi^2\mathcal{G}c\tau^3d}
\int_0^\infty\frac{\eta^2\,d\eta}{e^{\eta/y}-1}\,
\arctan\frac{\mathcal{G}\eta}x\nonumber\\
={}&{}\left\{\begin{array}{ll}
\displaystyle{\pi^2T^4}/({120\,\hbar^3c^2}), & y\ll{x}/\mathcal{G},\\
\displaystyle{\zeta(3)\,T^3}/({8\pi\hbar^2{c}\mathcal{G}d}), & y\gg{x}/\mathcal{G}.
\end{array}\right.
\label{eq:J1expand}
\end{align}
The ${u}$ integral converges at ${u}\sim1$ and $u\sim\sqrt{x/(\mathcal{G}y)}$ in the two cases.
In the first case, $y\ll{x}/\mathcal{G}$, the assumption $\mathcal{G}u\gg1$, as well as the condition to expand the exponential, ${u}{y}/x\ll1$, are satisfied automatically. In the second case, $y\gg{x}/\mathcal{G}$, both conditions translate into ${y}\ll\mathcal{G}{x}$.\\
(v) For $y\gg1$ but $y\ll\mathcal{G}{u}$ we still have $\eta\sim{y}$, so the denominator can be approximated as $4\mathcal{G}{u}^2\eta^2(1+\mathcal{G}{u}^2/x)^2$:
\begin{align}
J_\mathrm{TM}^\mathrm{t}={}&{}\frac{\hbar}{4\pi^2c^2\tau^4}
\int_0^\infty\frac{\eta\,d\eta}{e^{\eta/y}-1}
\int_0^1\frac{{u}\,d{u}}{[1+(\mathcal{G}/x){u}^2]^2}\nonumber\\
={}&{}\frac{T^2/\hbar}{48c\tau(c\tau+\mathcal{G}d)}.
\label{eq:J2expand}
\end{align}
Since the convergence occurs at ${u}\sim\min\{1,\sqrt{x/\mathcal{G}}\}$, $\eta\sim{y}$, the assumption $y\ll\mathcal{G}{u}$ is satisfied if ${y}\ll\min\{\sqrt{\mathcal{G}x},\mathcal{G}\}$; if so, the condition $uy/x\ll1$ to expand the exponential is satisfied automatically. Thus, Eq.~(\ref{eq:J2expand}) is valid when $1\ll{y}\ll\min\{\sqrt{\mathcal{G}x},\mathcal{G}\}$.\\
(vi) For $y\gg{1},\mathcal{G}{u}$, the Bose function is $y/\eta$, so we integrate over $\eta$ exactly (convergence at $\eta\sim1+2\mathcal{G}{u}$), and obtain
\begin{align}
J_\mathrm{TM}^\mathrm{t}={}&{}\frac{\mathcal{G}^2T}{2\pi{c}^2\tau^3}
\int_0^1\frac{{u}^3\,d{u}}{(1+2\mathcal{G}{u})[1+(\mathcal{G}/x){u}^2]^2}\nonumber\\
{}&{}=\left\{\begin{array}{ll}
Tx^{3/2}/(16{c}^2\tau^3\mathcal{G}^{1/2}), & {x}\ll\mathcal{G},\\
\mathcal{G}T/(12\pi{c}^2\tau^3), & x\gg\mathcal{G},
\end{array}\right.\label{eq:J3expand}
\end{align}
the convergence occurring at ${u}\sim\min\{\sqrt{x/\mathcal{G}},1\}$. At $x\ll\mathcal{G}$ the condition to expand $e^{i\xi/x}$ is not fulfilled, since we automatically have $\xi/x=u\eta/x\sim1$. At $x\gg\mathcal{G}$, we have $\mathcal{G}{u}\gg1$ automatically, while ${u}\eta/x\sim\mathcal{G}/x$, so the second expression Eq.~(\ref{eq:J3expand}) is valid at $x,y\gg\mathcal{G}$.

\begin{figure}
\includegraphics[width =0.45\textwidth]{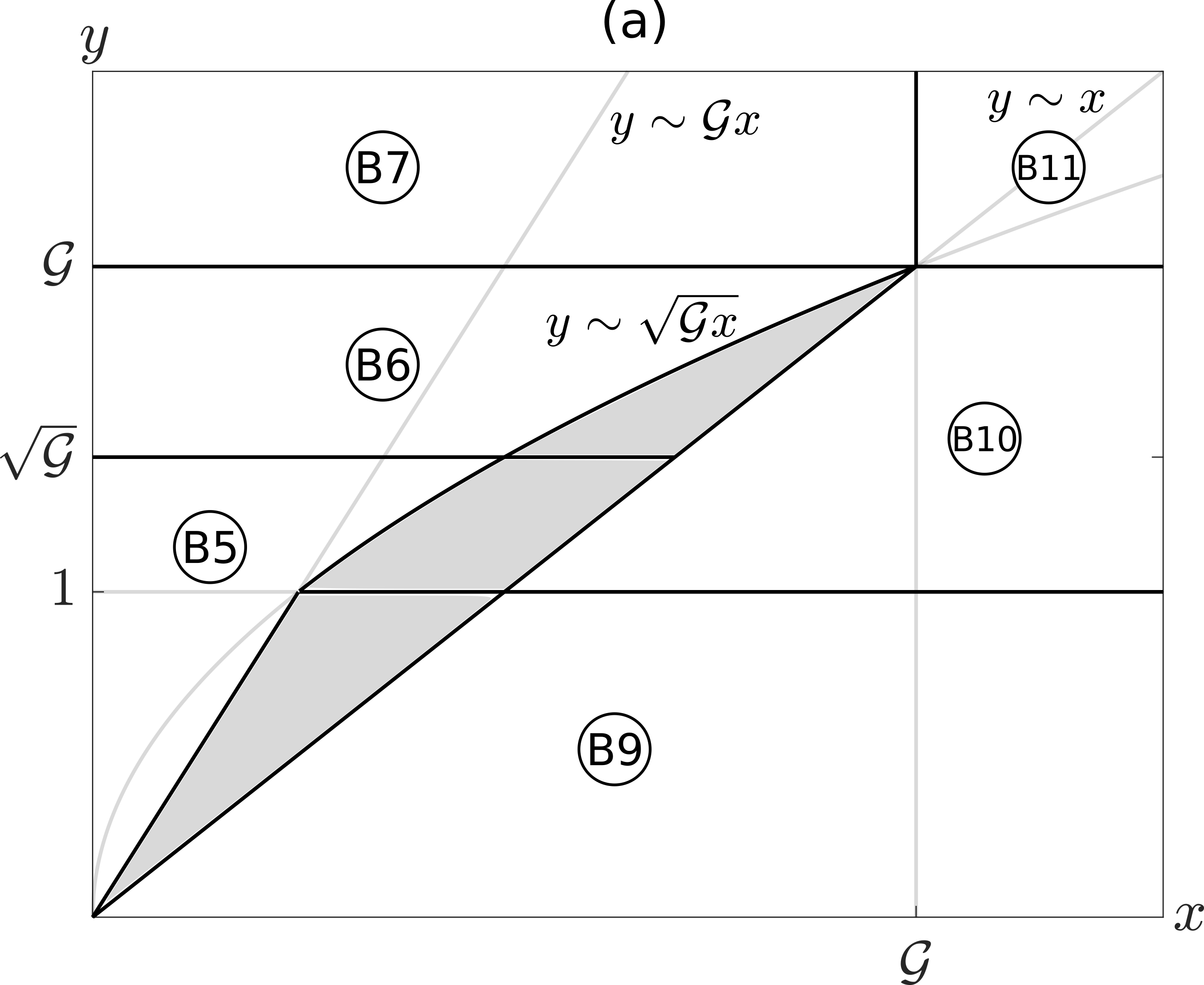}
\includegraphics[width =0.45\textwidth]{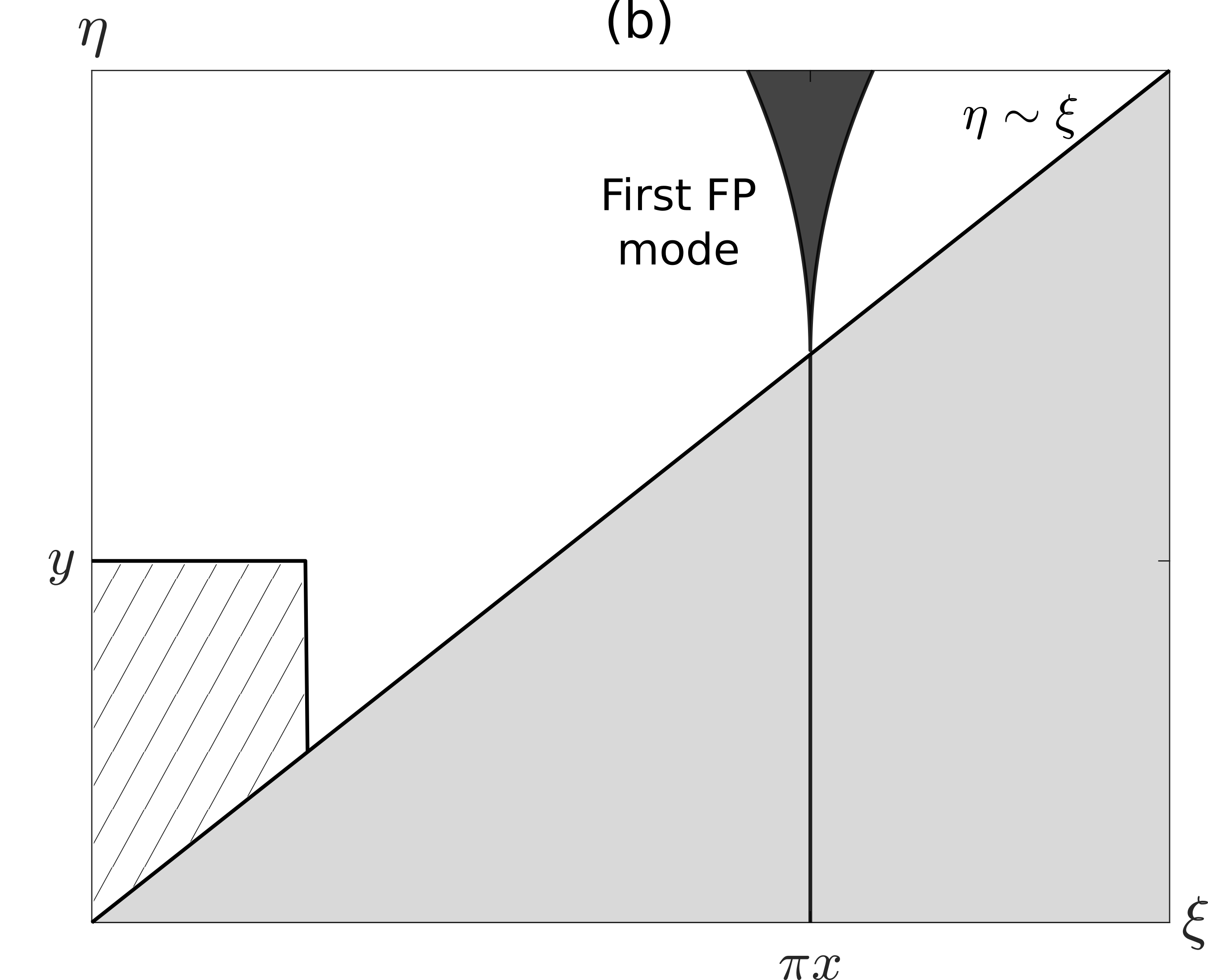}
\caption{(a)~Regions of validity of Eqs.~(\ref{eq:FP1=})--(\ref{eq:J3expand}) in the $(x,y)$ plane. In the shaded regions there are two valid contributions. (b)~The $(\xi,\eta)$ plane with the integration domain $\xi<\eta$ (the shaded area does not belong to the integration domain). The hatched area at $\xi\ll{x}$ contributes to Eqs.~(\ref{eq:J1expand}),~(\ref{eq:J2expand}).}
\label{fig:TMtrav}
\end{figure}

We schematically show the regions of validity of Eqs.~(\ref{eq:FP1=})--(\ref{eq:J3expand}) in the $(x,y)$ plane in Fig.~\ref{fig:TMtrav}(a). In the overlapping region at $y\gg{x}$ both $\xi\ll{x}$ and $\xi\gg{x}$ contributions are valid, but the Fabry-Perot contributions from $\xi\gg{x}$ naturally dominate. At $y\ll{x}$ the Fabry-Perot contributions are suppressed as $e^{-\pi{x}/y}=e^{-\pi\hbar{c}/(Td)}$, since the temperature is lower than the first Fabry-Perot mode energy $\pi\hbar{c}/d$. Nevertheless, it turns out that the prefactor in front of the exponential is large, so the contribution from the first mode (the one with the weakest exponential), coming from the narrow region around $\xi=\eta=\pi{x}$ [see Fig.~\ref{fig:TMtrav}(b)] should be included together with the contribution from $\xi\ll{x}$, as long as $x,y\ll\mathcal{G}$ (otherwise, the mode is overdamped because of low reflectivity).

To pick up the first Fabry-Perot mode contribution, we approximate the Bose function by $e^{-\eta/y}$ and set $\eta=\pi{x}$ everywhere else in the integrand, which is a smooth function of~$\eta$. We also set $\xi=\pi{x}$ everywhere in the integrand except the exponential $e^{i\xi/x}$ in $D^p_+$ [Eq.~(\ref{eqn:DpmTM=})]. Then we find the minimum of $D^p_+$ as a function of $\xi$, reached at $\xi_\mathrm{min}=\pi{x}(1-x/\mathcal{G})+O(x^3/\mathcal{G}^2)$, and approximate near the minimum
\begin{align}
D^p_+={}&{}(\pi\mathcal{G}x)^2\left|\frac{1-i\pi{x}}{\mathcal{G}}+1+e^{i\xi/x}\right|^2
\nonumber\\
\approx{}&{}(\pi{x})^2\left(1+\frac{\pi^2x^2}{2\mathcal{G}}\right)^2+(\pi\mathcal{G})^2(\xi-\xi_\mathrm{min})^2.
\end{align}
Then, the integration over $\pi{x}<\eta<\infty$ and $-\infty<\xi-\xi_\mathrm{min}<\infty$ gives
\begin{equation}\label{eqn:FPmode}
J_\mathrm{TM}^\mathrm{t}=\frac{\pi{c}T}{2d^3[2\mathcal{G}+(\pi{c}\tau/d)^2]}\,e^{-\pi\hbar{c}/(Td)}.
\end{equation}

\subsection{TE travelling contribution}
The TE travelling contribution to Eq.~(\ref{eqn:J(T)=}) can be rewritten exactly as
\begin{subequations}\begin{align}
&J_\mathrm{TE}^\mathrm{t}=\frac{\hbar\mathcal{G}^2}{\pi^2c^2\tau^4}
\int_0^\infty\frac{\eta^3\,d\eta}{e^{\eta/y}-1}\,
\int_0^\eta\frac{\xi^3\,d\xi}{D^s_+D^s_-},\label{eq:TEradiative}\\
&D^s_\pm\equiv|\xi(1-i\eta)+\mathcal{G}\eta(1\pm{e}^{i\xi/x})|^2.
\label{eqn:DpmTE=}
\end{align}\end{subequations}
For $\mathcal{G}\ll1$, we may not simply set $\mathcal{G}\to0$ in the denominator, as we did in the TM case: here this leads to a logarithmic divergence at $\xi\to0$. To see how the divergence is cut off, we note that convergence scale of the $\eta$~integral is the same as in the TM case: $\eta\sim{y}$ if $y\ll1$ and  $\eta\sim1$ if $y\gg1$. This gives the small-$\xi$ cutoff scales $\xi\sim\mathcal{G}\eta$ and $\xi\sim\mathcal{G}$, respectively. As a result,
\begin{align}
J_\mathrm{TE}^\mathrm{t}={}&{}\frac{\hbar\mathcal{G}^2}{\pi^2c^2\tau^4}\int_0^\infty
\frac{\eta^3\,d\eta}{(1+\eta^2)^2(e^{\eta/y}-1)}\ln\frac{\eta}{\min\{\mathcal{G},\mathcal{G}\eta\}} \nonumber \\
={}&{}
\left\{\begin{array}{ll}
[(\pi^2\mathcal{G}^2T^4)/({15\,\hbar^3c^2})]\ln(1/\mathcal{G}), & y\ll 1,\\
{}[(\mathcal{G}^2T)/(4\pi{c}^2\tau^3)]
\ln[(T\tau)/(\mathcal{G}\hbar)], & y\gg1,
\end{array}\right. 
\label{eq:TEtsmallG=}
\end{align}
The overall map of behaviours in parameter space is therefore equivalent to the TM travelling case given in Eq.~(\ref{eq:TMtsmallG=}), but the TE contribution~(\ref{eq:TEtsmallG=}) is always dominant due to the logarithmic factors. 

The calculation for $\mathcal{G}\gg1$ is very similar to that of the TM travelling wave contribution. Focusing firstly on the cases where $\xi\gg x$ so the exponentials $e^{i\xi/x}$ oscillate fast, the averaged contribution from Eq.~(\ref{eq:TEradiative}) via Eq.~(\ref{eq:fastoscillation}) is given by
\begin{align}
J_\mathrm{TE}^\mathrm{t}=\frac{\hbar\mathcal{G}^2}{\pi^2c^2\tau^4}\int_0^\infty\frac{\eta^3\,d\eta}{e^{\eta/y}-1}\int_0^\eta
\frac{\xi^2\,d\xi}{\xi\left(\eta^2+1\right)+2\mathcal{G}\eta} \nonumber \\
{}\times\frac{1}{\xi^2\left(\eta^2+1\right)+2\mathcal{G}\xi\eta+2\left(\mathcal{G}\eta\right)^2}.
\label{eqn:TEfastosc}
\end{align}
At low frequency the system Fabry-Perot modes are indicated, as in the TM case, in the minima in $D^s_{\pm}$, this time important when $\mathcal{G}\eta \gg \xi\sqrt{1+\eta^2}$. The integral in $\eta$ may again converge at $\eta\sim y$ due to the Bose function, or something else if $y$ is too large. We may identify the same regions as in the TM case. \\ 
(i) For $y\ll\sqrt{\mathcal{G}}$, we have that $\xi \sim \eta \sim y$, so the fast oscillation condition is $x\ll{y}$, and we may neglect all terms in the denominator containing $\xi$:
\begin{equation}
J_\mathrm{TM}^\mathrm{t}=\frac{\hbar\mathcal{G}^2}{\pi^2c^2\tau^4}
\int_0^\infty\frac{\eta^3\,d\eta}{e^{\eta/y}-1}\,
\int_0^\eta\frac{\xi^2\,d\xi}{4(\mathcal{G}\eta)^3}
=\frac{\pi^2T^4}{180\,\hbar^3c^2\mathcal{G}}.\label{eq:TEFP1=}
\end{equation}
(ii) For $\sqrt{\mathcal{G}}\ll{y}\ll\mathcal{G}$, we keep  $\xi\eta^2$ in the denominator in the first line of Eq.~(\ref{eqn:TEfastosc}) and $2\left(\mathcal{G}\eta\right)^2$ in the second line (again, oscillations are fast when $x\ll{y}$):
\begin{equation}
J_\mathrm{TE}^\mathrm{t}=\frac{\hbar\mathcal{G}^2}{\pi^2c^2\tau^4}
\int_0^\infty\frac{\eta^3\,d\eta}{e^{\eta/y}-1}\,
\int_0^\eta\frac{\xi^3\,d\xi}{2(\mathcal{G}\xi)^2\eta^4}
=\frac{T^2}{24\,\hbar{c}^2\tau^2}.\label{eq:TEFP2=}
\end{equation}
(iii) For $y\gg\mathcal{G}$, we expand the Bose function to give $y/\eta$ and retain $\xi\eta^2$ in the first line of Eq.~(\ref{eqn:TEfastosc}) and $\left(\xi\eta\right)^2 + 2\left(\mathcal{G}\eta\right)^2$ in the second. The integrals converge at $\xi,\,\eta\sim\mathcal{G}$ so the oscillations are fast when $x\ll\mathcal{G}$:
\begin{equation}
J_\mathrm{TE}^\mathrm{t}=\frac{\hbar\mathcal{G}^2}{\pi^2c^2\tau^4}
\int_0^\infty{y}\eta^2\,d\eta
\int_0^\eta\frac{\xi\,d\xi}{(\xi^2+2\mathcal{G}^2)\eta^4}=
\frac{\sqrt{2}\,\mathcal{G}T}{4\pi{c}^2\tau^3}.
\end{equation}
For the contributions coming from $\xi\ll x$, we expand the exponential $e^{i\xi/x}$:
\begin{align}
J_\mathrm{TE}^\mathrm{t}={}&{}\frac{\hbar\mathcal{G}^2}{\pi^2c^2\tau^4}\int_0^\infty
\frac{\eta^3\,d\eta}{e^{\eta/y}-1}\,\frac{1}{1+\eta^2(1+\mathcal{G}/x)^2}\nonumber\\
{}&{}\times\int_0^\eta\frac{\xi\,d\xi}{(\xi+2\mathcal{G}\eta)^2+\xi^2\eta^2}.
\end{align}
Since $\xi<\eta$ and $\mathcal{G}\gg1$ we may neglect $\xi$ in the first bracket of the denominator in the last line. This allows the simple integration over $\xi$:
\begin{align}
J_\mathrm{TE}^\mathrm{t}={}&{}\frac{\hbar\mathcal{G}^2}{2\pi^2c^2\tau^4}\int_0^\infty
\frac{\eta\,d\eta}{e^{\eta/y}-1}\,
\frac{\ln[1+\eta^2/(2\mathcal{G})^2]}{1+\eta^2(1+\mathcal{G}/x)^2},
\end{align}
the condition for the expansion of the exponential $e^{i\xi/x}$ becoming $\eta\ll{x}$.

\begin{figure}
\includegraphics[width =0.45\textwidth]{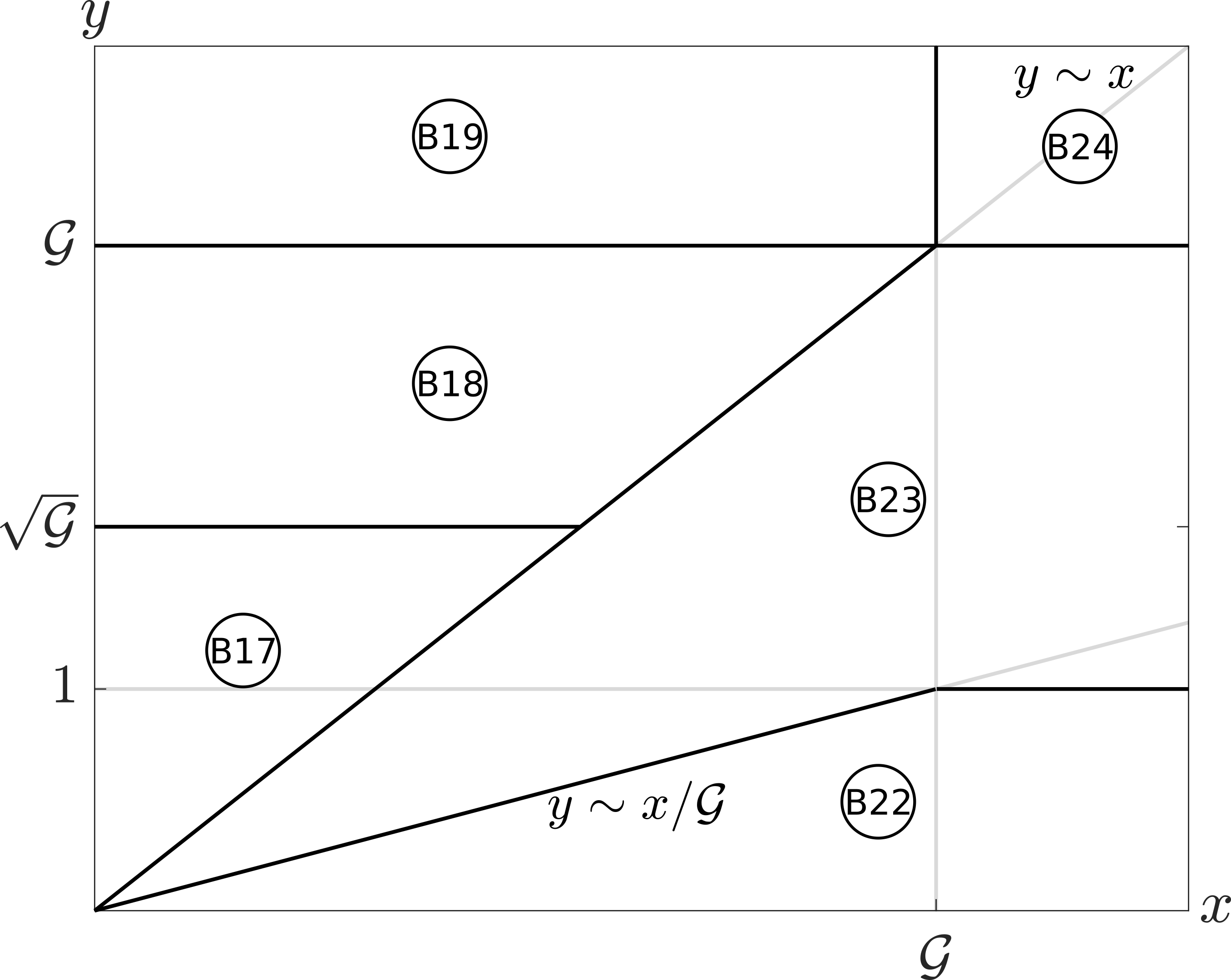}
\caption{Regions of validity of Eqs.~(\ref{eq:TEFP1=})--(\ref{eq:J3expandTE}) in the $(x,y)$ plane.}
\label{fig:TEtrav}
\end{figure}

There are three possible cutoff scales for $\eta$: $y$, $\mathcal{G}$, and $(1+\mathcal{G}/x)^{-1}$. Which one of the three is effective, depends on the positioning of~$y$ with repect to other scales. Again, three cases arise.\\
(iv) For $y\ll (1+\mathcal{G}/x)^{-1}<1$, the logarithm is expanded for small argument and the second term in the denominator is neglected since the integral converges at $\eta\sim y$.
The condition $y\ll x$ for the expansion of $e^{i\xi/x}$ is satisfied automatically:
\begin{equation}
J_\mathrm{TE}^\mathrm{t}=\frac{\hbar}{8\pi^2c^2\tau^4}
\int_0^\infty\frac{\eta^3\,d\eta}{e^{\eta/y}-1} = \frac{\pi^2 T^4}{120\hbar^3 c^2}.
\label{eq:J1expandTE}
\end{equation}
(v) For $(1+\mathcal{G}/x)^{-1}\ll y\ll\mathcal{G}$, the integral is still determined by $\eta\sim y$, but the second term in the denominator dominates. $e^{i\xi/x}$ may be expanded when $y\ll x$: 
\begin{equation}
J_\mathrm{TE}^\mathrm{t}=\frac{\hbar}{8\pi^2c^2\tau^4}\,
\frac{x^2}{(x+\mathcal{G})^2}
\int_0^\infty\frac{\eta^3\,d\eta}{e^{\eta/y}-1} = \frac{T^2/\hbar}{48\left(c\tau+\mathcal{G}d\right)^2}.
\label{eq:J2expandTE}
\end{equation}
(vi) For $y\gg \mathcal{G}$, the Bose function is $y/\eta$ and the integral converges at $\eta\sim\mathcal{G}$ so we retain the logarithm, and $e^{i\xi/x}$  may be expanded as long as $\mathcal{G}\ll x$:
\begin{equation}
J_\mathrm{TE}^\mathrm{t}=\frac{\mathcal{G}^2T\tau}{2\pi^2c^2\tau^4}\,
\frac{x^2}{(x+\mathcal{G})^2} \int\limits_0^\infty \frac{d\eta}{\eta^2} \ln\left(1+\frac{\eta^2}{4\mathcal{G}^2}\right) 
=\frac{\mathcal{G}T}{4\pi c^2\tau^3}.
\label{eq:J3expandTE}
\end{equation}
We schematically show the regions of validity of Eqs.~(\ref{eq:TEFP1=})--(\ref{eq:J3expandTE}) for $\mathcal{G}\gg1$ in the $(x,y)$ plane in Fig.~\ref{fig:TEtrav}, where there is no such overlap as in the TM case Fig.~\ref{fig:TMtrav}(a).

As in the TM case, the first Fabry-Perot mode contribution should be included together with the $\xi\ll x$ contributions as long as $x,y\ll\mathcal{G}$. The same procedure is performed whereby the minimum of $D_+^s$ [Eq.~(\ref{eqn:DpmTE=})] near $\xi=\pi x$ is found, allowing the integrand to be approximated by a Lorentzian. The minimum and therefore the eventual contribution is found to be identical to the TM case, Eq.~(\ref{eqn:FPmode}).

\subsection{TM evanescent contribution}

\begin{figure}
\includegraphics[width =0.45\textwidth]{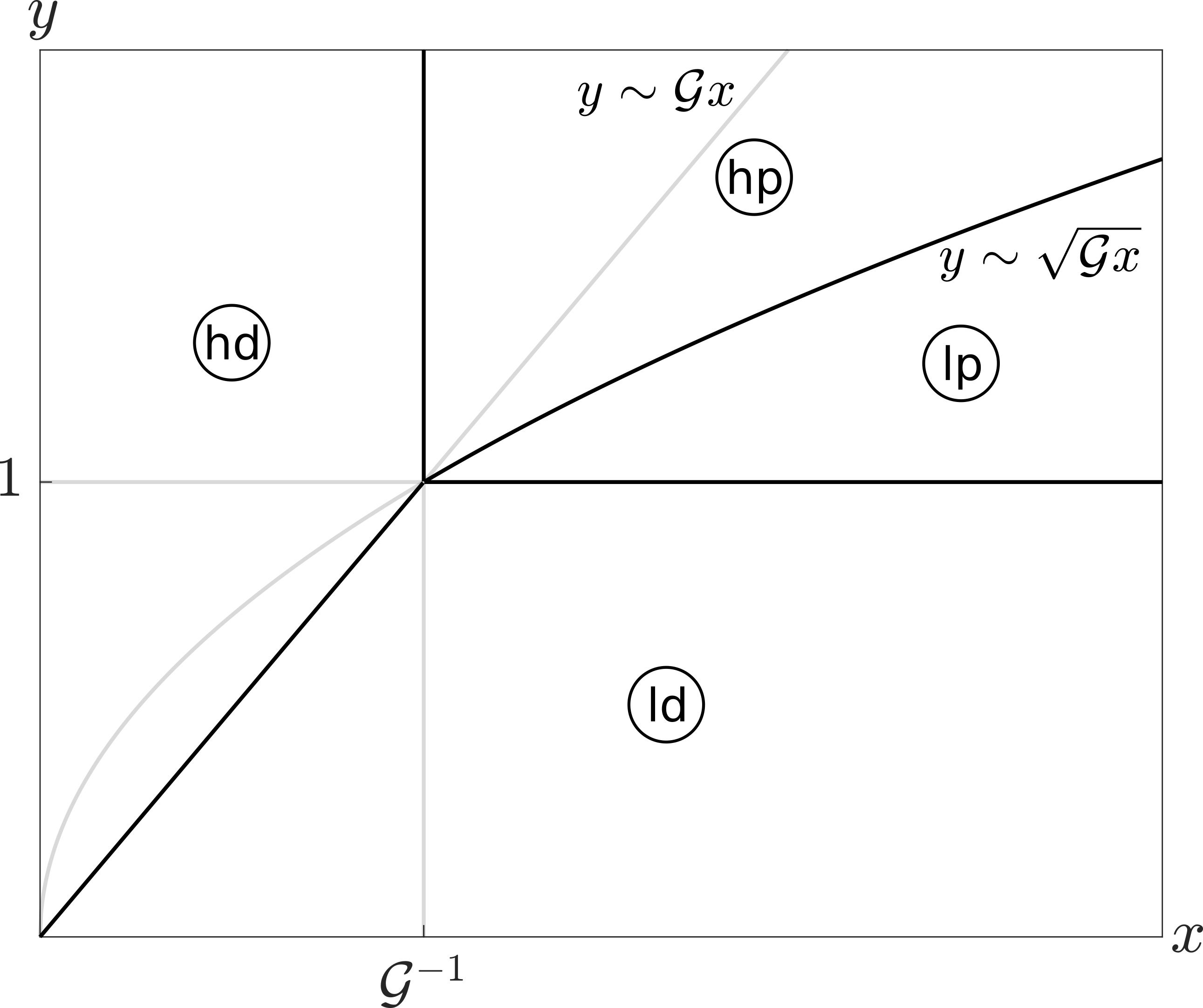}
\caption{Regions of validity of Eqs.~(\ref{eq:TMevld})--(\ref{eq:TMevhp}) in the $(x,y)$ plane.}
\label{fig:TMev}
\end{figure}

The TM evanescent contribution to Eq.~(\ref{eqn:J(T)=}) can be rewritten exactly as
\begin{subequations}\begin{align}
&J_\mathrm{TM}^\mathrm{e}=\frac{\hbar\mathcal{G}^2}{\pi^2c^2\tau^4}
\int_0^\infty\frac{ \eta^3\,d\eta}{e^{\eta/y}-1}\,
\int_0^\infty\frac{e^{-2\xi/x} \xi^3 \,d\xi}{\tilde{D}^p_+\tilde{D}^p_-},\label{eq:TMevanescent}\\
&\tilde{D}^p_\pm\equiv|\eta(1-i\eta)+i\mathcal{G}\xi(1\pm{e}^{-\xi/x})|^2.
\label{eqn:DepmTM=}
\end{align}\end{subequations}
This integral turns out to be exactly identical to that already calculated in Ref.~\cite{Wise2020} when the spatial dispersion of the conductivity is neglected (namely, Eqs.~(1), (10) and (11) of Ref.~\cite{Wise2020}). That is to say that in the present system the Coulomb limit ($c\rightarrow\infty$) amounts to taking only the exact TM evanescent contributions to the heat current, while neglecting the rest. 
Thus, we can simply rewrite the results of Ref.~\cite{Wise2020} in terms of~$\mathcal{G}$:
\begin{subequations}\begin{align}
J_\mathrm{ld} ={}&{} \frac{\zeta(3) T^3}{8\pi\hbar^2 c \mathcal{G} d}, \label{eq:TMevld} \\ 
J_\mathrm{hd} ={}&{} \frac{c\mathcal{G} T}{16\pi d^3 } \label{eq:TMevhd} \\ 
J_\mathrm{lp} ={}&{} \frac{\zeta(3) T^3}{4\pi\hbar^2 c \mathcal{G} d}, \label{eq:TMevlp} \\ 
J_\mathrm{hp} ={}&{} \frac{T}{16\pi\tau d^2} \mathcal{L}(\mathcal{G}c\tau/d), \label{eq:TMevhp} 
\end{align}\end{subequations}
where $\mathcal{L}(x)$ is a slow logarithmic function approximately given by:
\begin{equation}
\mathcal{L}(x)\approx\frac{4\ln^3x}{1+(\ln{x})/\ln(1+\ln{x})}.
\end{equation}
The domains of validity of the contributions are shown in Fig.~\ref{fig:TMev}. Note that expression~(\ref{eq:TMevld}) equals the travelling contribution~(\ref{eq:J1expand}).

\subsection{TE evanescent contribution}
The TE evanescent contribution to Eq.~(\ref{eqn:J(T)=}) can be rewritten exactly as
\begin{subequations}\begin{align}
&J_\mathrm{TE}^\mathrm{e}=\frac{\hbar\mathcal{G}^2}{\pi^2c^2\tau^4}
\int_0^\infty\frac{\eta^3\,d\eta}{e^{\eta/y}-1}\,
\int_0^\infty \frac{e^{-2\xi/x} \xi^3\,d\xi}{\tilde{D}^s_+\tilde{D}^s_-},\label{eq:TEevanescent}\\
&\tilde{D}^s_\pm\equiv|i\xi(1-i\eta)+\mathcal{G}\eta(1\pm{e}^{-\xi/x})|^2.
\label{eqn:DepmTE=}
\end{align}\end{subequations}
Despite the apparent similarity to the corresponding TE travelling contribution Eq.~(\ref{eq:TEradiative}), there is no longer oscillatory behaviour in the denominator, so the resulting contributions are completely different. In $\eta$ there are two possible decay scales: $\eta\sim y$ from the Bose function, and $\eta \sim \xi/(\mathcal{G}+\xi)$ from $\tilde{D}^s_+\tilde{D}^s_-$. 

\begin{figure}
\includegraphics[width =0.45\textwidth]{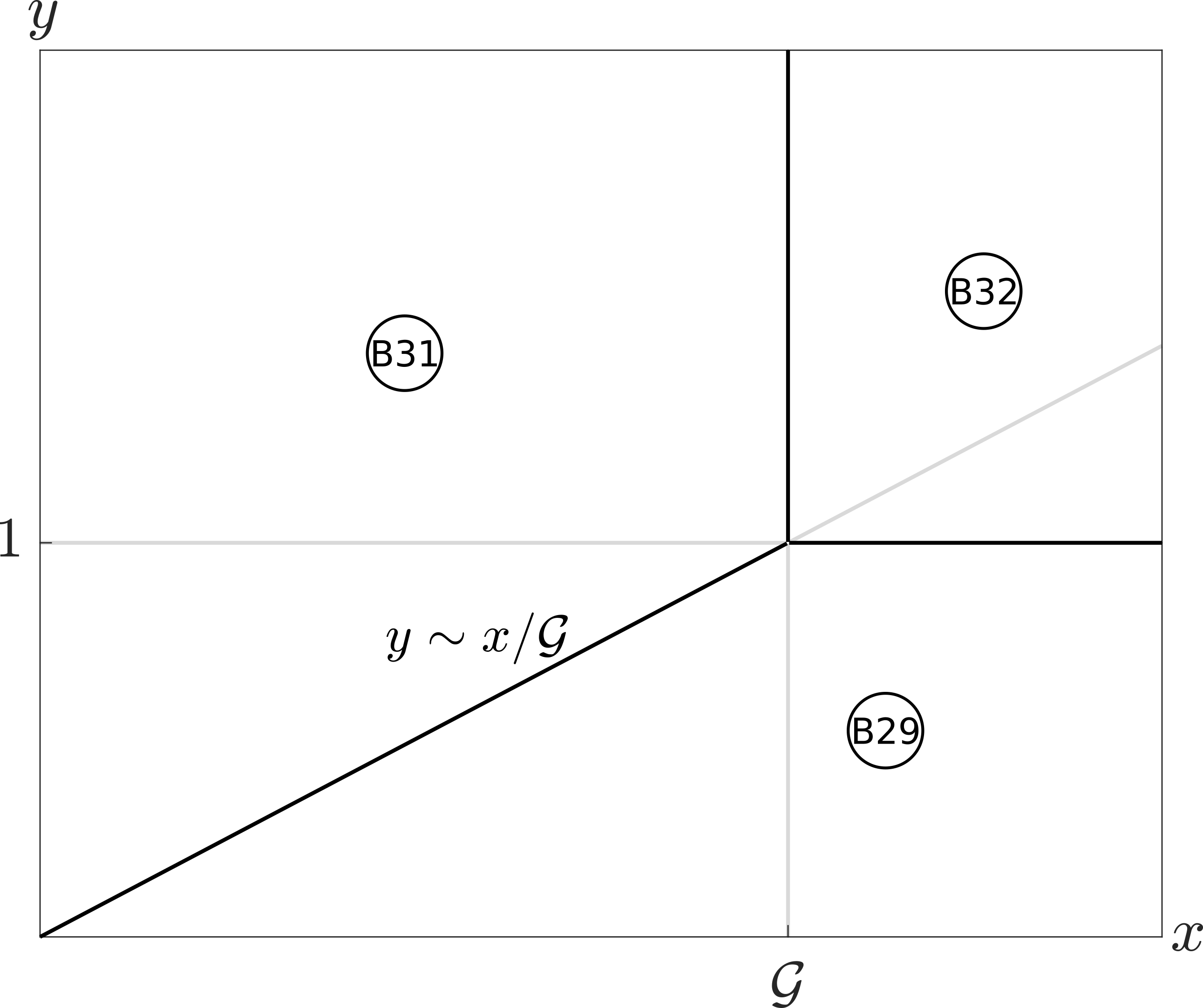}
\caption{Regions of validity of Eqs.~(\ref{eq:TEe1=})--(\ref{eq:TEe3=}) in the $(x,y)$ plane.}
\label{fig:TEev}
\end{figure}

In the low temperature case $y\ll \xi/(\mathcal{G}+\xi) <1$ where the temperature cutoff is effective, expanding $e^{-\xi/x}\approx 1$ leads to logarithmic divergence at $\xi\rightarrow\infty$. The large $\xi$ cutoff scale is therefore given by the decay scale of the exponential, $\xi\sim x$, leading to the result [valid for $y\ll x/(\mathcal{G}+x)$]:
\begin{equation}\label{eq:TEe1=}
J_\mathrm{TE}^\mathrm{e} = \frac{\hbar\mathcal{G}^2}{\pi^2c^2\tau^4}\int\limits_0^\infty \frac{\eta^3 \,d\eta}{e^{\eta/y}-1} \ln \frac{x}{\mathcal{G} \eta} = \frac{\pi^2 \mathcal{G}^2 T^4}{15 \hbar^3 c^2} \ln\frac{\hbar c}{\mathcal{G}Td}.
\end{equation}
For high temperatures $y\gg\xi/(\mathcal{G}+\xi)$ the Bose function is $y/\eta$ and it is convenient to perform integration over $\eta$ first keeping $\tilde{D}_\pm^s$ exact:
\begin{equation}\label{eq:TEehT}
J_\mathrm{TE}^\mathrm{e} = \frac{\mathcal{G}^2 T}{4 \pi c^2\tau^3}\int\limits_0^\infty \frac{e^{-2\xi/x} \xi^2 \,d\xi}{(\mathcal{G}+\xi) [ (\mathcal{G}+\xi)^2-\mathcal{G}^2e^{-2\xi/x})]},
\end{equation}
where the integrand may decay due to the exponential or the denominator. If $x\ll\mathcal{G}$ the exponential is clearly active and terms in $\xi$ may be neglected in the denominator (the expansion of the Bose function is valid for $y\gg x/\mathcal{G}$): 
\begin{equation}
J_\mathrm{TE}^\mathrm{e} = \frac{T}{4\pi^2 \mathcal{G} c^2\tau^3} \int_0^\infty \frac{\xi^2 \, d\xi }{e^{2\xi/x}-1}  = \frac{\zeta(3) c T}{16\pi \mathcal{G}d^3}.
\label{eq:TEe2=}
\end{equation}
If $x\gg\mathcal{G}$, expansion of $e^{-\xi/x}\approx 1$ in Eq.~(\ref{eq:TEehT}) again leads to logarithmic divergence at $\xi\rightarrow \infty$. As in the low temperature case, the divergence is cut off by $\xi\sim x$ (the expansion of the Bose function is valid for $y\gg 1$): 
\begin{equation}
J_\mathrm{TE}^\mathrm{e} =  \frac{\mathcal{G}^2 T}{4\pi c^2\tau^3} \int_0^{\sim x} \frac{\xi \, d\xi }{(\xi+\mathcal{G})(\xi+2\mathcal{G})}  = \frac{\mathcal{G}^2 T}{4\pi c^2\tau^3} \ln \frac{c\tau}{\mathcal{G}d} .
\label{eq:TEe3=}
\end{equation}
The domains of validity of the TE evanescent contributions are shown in Fig.~\ref{fig:TEev}.

\section{Heat current between three-dimensional metallic half-spaces}
\label{app:3D}

In this section we give a derivation of asymptotic expressions for the heat current between two three-dimensional semi-infinite metallic half-spaces, separated by a vacuum gap~$d$, essentially reproducing the results obtained in Ref.~\cite{Polder1971}. We take two identical metals, described by the complex dielectric functions $\varepsilon(\omega) = 1+4\pi i\sigma_\text{3D}/\omega$, where $\sigma_\text{3D}$ is the three-dimensional dc conductivity, which can be written in terms of the bulk plasma frequency $\omega_p$ and the electron relaxation time~$\tau$ as $4\pi\sigma_\text{3D}=\omega_p^2\tau$, and assumed to be temperature-independent. For conventional metals, $4\pi\sigma_\text{3D}\gg\omega_p\gg1/\tau$, and it is natural to assume $T\ll\hbar/\tau$ (indeed, $\tau=10^{-14}\:\mbox{s}$ corresponds to 760~K), so that for all relevant frequencies $\varepsilon(\omega)\approx4\pi{i}\sigma_{3D}/\omega\gg1$. Focusing on the local response regime, we assume to be in the normal skin effect regime, characterised by the frequency-dependent skin depth $\delta_\omega$ and its value at $\omega=T/\hbar$:
\begin{equation}\label{eq:skin}
\delta_\omega=\frac{c}{\sqrt{2\pi\sigma_{3D}\omega}},\quad
\delta_T\equiv\frac{c}{\sqrt{2\pi\sigma_{3D}T/\hbar}}
\end{equation}

Since the metals are semi-infinite there can be no transmitted radiation and therefore the Joule losses are equal unambiguously to the average Poynting vector in the gap. 
The heat current per unit area $J(T)$ may once again be written in the form of Eq.~(\ref{eqn:J(T)=}), but without the third term in the emissivities in Eq.~(\ref{eqn:emissreflect}) (corresponding to transmission in the two-dimensional case), and with the reflectivities being just the Fresnel coefficients \cite{Volokitin2001}:
\begin{equation}
r_p = \frac{ q_z - q_z'/\varepsilon}{ q_z + q_z'/\varepsilon},\quad
r_s =\frac{q_z-q_z'}{q_z+q_z'},
\end{equation}
where $q_z' = \sqrt{[\varepsilon(\omega)-1](\omega^2/c^2)+q_z^2}$ is the normal component of the complex wavevector describing the electric and magnetic fields inside the metal, while $q_z$~is the same in the vacuum gap. As in the two-dimensional case, the contributions from travelling and evanescent waves for each polarisation are computed separately. 

\subsection{TM travelling contribution}
The contribution may be written exactly as
\begin{equation} \label{eq:3dJ_TMt=}
J_\mathrm{TM}^\mathrm{t} =  \frac{\hbar}{4\pi^2}\int_0^\infty \frac{\omega \,d\omega}{e^{\hbar\omega/T}-1} \int_0^{\omega/c} q_z \, dq_z\, \frac{\left(1-\left|r_p\right|^2\right)^2}{\left|1-r_p^2 e^{-2iq_zd}\right|^2}. 
\end{equation}
The Fresnel coefficient is simplified drastically by noticing that since $cq_z<\omega\ll 4\pi\sigma_\mathrm{3D}$ we may write
\begin{equation}
q_z'\approx \frac{\omega}{c}\sqrt{\varepsilon(\omega)},\quad
r_p\approx{1}-\frac{\omega}{cq_z}\,\frac2{\sqrt{\varepsilon(\omega)}}.
\label{eq:qz'}
\end{equation}
Focussing firstly on the case where the exponential in the denominator is oscillating fast, we may perform the same averaging according to Eq.~(\ref{eq:fastoscillation}), valid for $q_z\gg 1/d$, which translates into $d \gg \lambdabar_T$. This gives
\begin{align}
J_\mathrm{TM}^\mathrm{t} ={}&{}  \frac{\hbar}{4\pi^2c^2}\int_0^\infty \frac{\hbar\omega^3 \,d\omega}{e^{\hbar\omega/T}-1} 
\sqrt{\frac{\omega}{2\pi\sigma_\mathrm{3D}}} \nonumber  \\ 
={}&{}\frac{105\, \zeta(9/2)}{64\pi^{3/2}} \frac{\hbar(T/\hbar)^{9/2}}{\sqrt{2\pi\sigma_\mathrm{3D}} c^2}. 
\label{eq:3dJ_TMt1}
\end{align}
When $q_z\ll 1/d$, the exponential in the denominator of Eq.~(\ref{eq:3dJ_TMt=}) may be expanded as $1-2iq_zd$, so the denominator is approximately
\begin{equation}
1-r_p^2e^{-2iq_zd}\approx2iq_z\left[d-(1+i)\delta_\omega\,\frac{\omega^2}{c^2q_z^2}\right].
\end{equation}
This results in two expressions, depending on the relation between $d$ and the thermal skin depth $\delta_T$:
\begin{subequations}\begin{align}
J_\mathrm{TM}^\mathrm{t}= {}&{} \frac{\hbar}{4\pi^2}\int_0^\infty
\frac{\omega \,d\omega}{e^{\hbar\omega/T}-1} \int_0^{\omega/c}
\frac{\delta_\omega^2q_z\,dq_z\,}{[(cq_z/\omega)^2d-\delta_\omega]^2+\delta_\omega^2}\nonumber\\
= {}&{} \frac{15\,\zeta(7/2)}{128\sqrt\pi}\,
\frac{\hbar(T/\hbar)^{7/2}}{\sqrt{2\pi\sigma_{3D}}cd},\quad
d\gg\delta_T,\label{eq:3dJ_TMt2a}\\
= {}&{} \frac{\pi^2\hbar(T/\hbar)^4}{240\,c^2},\quad d\ll\delta_T.
\label{eq:3dJ_TMt2b}
\end{align}\end{subequations}
Note that in the first case the $q_z$ integral converges at $q_z\sim(\omega/c)\sqrt{\delta_\omega/d}\ll\omega/c$, so the expansion of $e^{-2iq_zd}$ is valid at $d\ll\lambdabar_T^2/\delta_T$, which is a weaker condition than $d\ll\delta_T$; this means that the small~$q_z$ contribution may coexist with that of Fabry-Perot modes, but it is subdominant.
In the second case $d\ll\delta_T$, the convergence is at $q_z\sim\omega/c$, so the condition $q_zd\ll1$ is automatically satisfied when $d\ll\delta_T$.

\subsection{TE travelling contribution}
The situation is quite analogous to the TM case.
The TE contribution is given by Eq.~(\ref{eq:3dJ_TMt=}), but with the replacement $r_p\to{r}_s$.
Instead of Eq.~(\ref{eq:qz'}), we have
\begin{equation}\label{eq:3dJ_TErsapprox}
q_z'\approx \frac{\omega}{c}\sqrt{\varepsilon(\omega)},\quad
r_s\approx-{1}+(1-i)q_z\delta_\omega.
\end{equation}
In the case of fast oscillation at $d\gg\lambdabar_T$, the denominator is again averaged using Eq.~(\ref{eq:fastoscillation}), leading to an expression, smaller than Eq.~(\ref{eq:3dJ_TMt1}) by a factor of~3.

When $d\ll\lambdabar_T$, expanding  $e^{-2iq_zd}\approx1-2iq_zd$, we obtain $1-r_s^2e^{-2iq_zd}\approx2iq_z[d-(1+i)\,\delta_\omega]$, which again results in two expressions:
\begin{subequations}\begin{align}
J_\mathrm{TE}^\mathrm{t} = {}&{} \frac{\hbar}{8\pi^2c^2}\int_0^\infty \frac{\omega^3 \,d\omega}{e^{\hbar\omega/T}-1}\,\frac{\delta_\omega^2(\omega)}{|d-(1+i)\delta_\omega|^2}\nonumber\\
= {}&{} \frac{\zeta(3)}{4\pi^2}\,\frac{\hbar(T/\hbar)^3}{2\pi\sigma_{3D}d^2},\quad d\gg\delta_T,\label{eq:3dJ_TEt2a}\\
= {}&{} \frac{\pi^2\hbar(T/\hbar)^4}{240\,c^2},\quad d\ll\delta_T.
\label{eq:3dJ_TEt2b}
\end{align}\end{subequations}
In contrast to the previous TM case, the integral is always dominated by $q_z\sim\omega/c$. In this case, analogously to Eq.~(\ref{eqn:FPmode}), one can also take into account the contribution of the first Fabry-Perot mode:
\begin{equation}
J_\mathrm{TE}^\mathrm{t} = \frac{\pi^2}4\,\frac{cT\delta_1}{d^4}\,
e^{\pi\hbar{c}/(Td)},
\end{equation}
where $\delta_1$ is $\delta_\omega$ corresponding to $\omega=\pi{c}/d$. This expression has an exponential smallness, but its prefactor is parametrically larger than Eq.~(\ref{eq:3dJ_TEt2a}).

\subsection{TM evanescent contribution}
The contribution may be written exactly as
\begin{equation} \label{eq:3dJ_TMe=}
J_\mathrm{TM}^\mathrm{e} =  \frac{\hbar}{\pi^2}\int_0^\infty \frac{\omega \,d\omega}{e^{\hbar\omega/T}-1} \int_0^{\infty} \tilde{q}_z \, d\tilde{q}_z\, \frac{\left(\Im r_p \right)^2e^{-2\tilde{q}_zd}}{\left|1-r_p^2 e^{-2\tilde{q}_zd}\right|^2}, 
\end{equation}
where the real integration variable $\tilde{q}_z$ is introduced since $q_z=i\tilde{q}_z$ is purely imaginary. Then $q_z' = \sqrt{\varepsilon(\omega)\,(\omega/c)^2-\tilde{q}_z^2}$. 

Let us first consider the case where the $\tilde{q}_z^2$ dominates over $\varepsilon(\omega/c)^2$ in the square root, that is $\tilde{q}_z\delta_\omega\gg1$. Then
\begin{equation}
r_p\approx1-\frac2{\varepsilon(\omega)}=1+\frac{i\omega}{2\pi\sigma_{3D}},
\end{equation}
and 
\begin{align}
J_\mathrm{TM}^\mathrm{e} = {}&{} \frac{\hbar}{4\pi^2}\int_0^\infty \frac{\omega \,d\omega}{e^{\hbar\omega/T}-1} 
\int_0^{\infty}
\frac{[\omega/(2\pi\sigma_{3D})]^2\,\tilde{q}_z \, d\tilde{q}_z}{\sinh^2\tilde{q}_zd+[\omega/(\pi\sigma_{3D})]^2} \nonumber\\ ={}&{}\frac{\pi^2}{60}\,\frac{\hbar(T/\hbar)^4}{(2\pi\sigma_{3D})^2d^2}
\ln\min\left\{\frac{2\pi\sigma_{3D}}{T/\hbar},\frac{\delta_T^2}{d^2}\right\}.\label{eq:3dJ_TMe1}
\end{align}
The $\tilde{q}_z$ integral is logarithmic, and is determined by a broad interval of $\tilde{q}_z$ from the upper cutoff $\sim1/d$ down to the lower cutoff: for $d\ll{c}/(2\pi\sigma_{3D})$ it is $(1/d)\sqrt{\omega/(2\pi\sigma_{3D})}$, while at larger distances the small~$\tilde{q}_z$ cutoff is determined by the condition of $\tilde{q}_z\delta_\omega\gg1$. The logarithmic region exists at all if $\varepsilon(\omega/c)^2$ can be neglected at $\tilde{q}_z\sim1/d$, which translates into $d\ll\delta_T$.

In the opposite case, where we neglect $\tilde{q}_z\ll1/\delta_\omega$ in $q_z'$, we still assume $\tilde{q}_z\gg(\omega/c)/\sqrt{\varepsilon}\sim(\omega/c)^2\delta_\omega$, so the reflection coefficient is still close to unity:
\begin{equation}
r_p \approx 1 + (1+i)\, \frac{\omega}{c\tilde{q}_z}\sqrt{\frac{\omega}{2\pi\sigma_\mathrm{3D}}}.
\end{equation}
Then the $\tilde{q}_z$ integral is determined by small $\tilde{q}_z\sim\sqrt{\omega/(cd)}[\omega/(2\pi\sigma_{3D})]^{1/4}\ll1/d$, so $\sinh\tilde{q}_zd\approx\tilde{q}_zd$:
\begin{align}
J_\mathrm{TM}^\mathrm{e} = {}&{} \frac{\hbar}{4\pi^2}\int_0^\infty \frac{\omega \,d\omega}{e^{\hbar\omega/T}-1} \nonumber\\ {}&{}\times\int_0^{\infty}\tilde{q}_z \, d\tilde{q}_z
\left[\left(\frac{\tilde{q}_z^2cd}\omega\sqrt{\frac{2\pi\sigma_{3D}}\omega}-1\right)^2+1\right]^{-1} \nonumber\\ 
={}&{}\frac{45\,\zeta(7/2) }{256 \sqrt\pi }\,\frac{\hbar(T/\hbar)^{7/2}}{\sqrt{2\pi\sigma_\mathrm{3D}}cd}.\label{eq:3dJ_TMe2}
\end{align}
The conditions $\tilde{q}_zd\ll1$, $(\omega/c)^2\delta_\omega\ll\tilde{q}_z\ll1/\delta_\omega$ result in the requirement
\begin{equation}\label{eq:3dJ_TMe_ineq}
\frac{T/\hbar}{2\pi\sigma_{3D}}\,\delta_T\ll{d}\ll
\frac{2\pi\sigma_{3D}}{T/\hbar}\,\delta_T.
\end{equation}
Note that the lower limit on~$d$ is smaller than $\delta_T$, so there is an interval where Eqs.~(\ref{eq:3dJ_TMe1}) and (\ref{eq:3dJ_TMe2}) are both valid, representing contributions from different regions of $\tilde{q}_z$ integration. However, when inequalities~(\ref{eq:3dJ_TMe_ineq}) hold, Eq.~(\ref{eq:3dJ_TMe2}) automatically dominates over Eq.~(\ref{eq:3dJ_TMe1}). Going to longer distances, where the assumption $(\omega/c)/\sqrt\varepsilon\ll\tilde{q}_z$ is violated, is not necessary since at such distances (well exceeding $\lambdabar_T$) the travelling wave contributions dominate; indeed, Eq.~(\ref{eq:3dJ_TMt1}) exceeds Eq.~(\ref{eq:3dJ_TMe2}) in the common wisdom region $d\gg\lambdabar_T$.

\subsection{TE evanescent contribution}
The TE contribution is given by Eq.~(\ref{eq:3dJ_TMe=}), but with the replacement $r_p\to{r}_s$. If we try to proceed as in the TM case and assume first $\tilde{q}_z\delta_\omega\gg1$, we obtain an integral diverging at small $\tilde{q}_z$, invalidating the assumption.

Making the opposite assumption, $q_z\delta_\omega\ll1$, for the reflection coefficient we obtain the same approximation~(\ref{eq:3dJ_TErsapprox}) with $q_z=i\tilde{q}_z$, which leads to
\begin{equation}
J_\mathrm{TE}^\mathrm{e} =  \frac{\hbar}{4\pi^2}\int\limits_0^\infty \frac{\omega \,d\omega}{e^{\hbar\omega/T}-1} \int\limits_0^{\infty} \frac{\tilde{q}_z^3 \, d\tilde{q}_z}%
{[(\sinh\tilde{q}_zd)/\delta_\omega+\tilde{q}_z]^2+\tilde{q}_z^2}.
\end{equation}
At $d\gg\delta_\omega$ the $\tilde{q}_z$ integral converges at $\tilde{q}_z\sim{1}/d$, and the resulting logarithmic $\omega$~integral
\begin{align}
J_\mathrm{TE}^\mathrm{e}={}&{}\frac{3\,\zeta(3)}{8\pi^2d^4}
\int\limits_0^\infty \frac{\delta_\omega^2\,\hbar\omega \,d\omega}{e^{\hbar\omega/T}-1}
=\frac{3\,\zeta(3)}{4\pi^2}\,\frac{c^2T}{2\pi\sigma_{3D}d^4}\ln\frac{d}{\delta_T},
\label{eq:3dJ_TEe2}
\end{align}
is cut off at low frequencies by the condition $d\sim\delta_\omega$, so that the validity condition is $d\gg\delta_T$.

For $d\ll\delta_T$ we are forced to consider $\tilde{q}_z\delta_\omega\sim1$ and use the exact expression $q_z'=\sqrt{2i-\tilde{q}_z^2\delta^2_\omega}/\delta_\omega$; however, we can safely set $d\to0$ as $1\pm{r}_s\sim{1}$. Then we obtain
\begin{equation}
\frac{\left(\Im r_s \right)^2}{\left|1-r_s^2\right|^2}
=\frac{(\Re{q}_z')^2}{4|q_z'|^2}=
\frac{1/2}{4+\tilde{q}_z^4\delta_\omega^4+\tilde{q}_z^2\delta_\omega^2\sqrt{4+\tilde{q}_z^4\delta_\omega^4}},
\end{equation}
which gives
\begin{equation}
J_\mathrm{TE}^\mathrm{e} =  \frac{2\pi\sigma_{3D}\hbar}{8\pi^2c}\int_0^\infty \frac{\omega^2 \,d\omega}{e^{\hbar\omega/T}-1} = \frac{\zeta(3)}{4\pi^2} \frac{2\pi\sigma_\mathrm{3D} \hbar(T/\hbar)^3}{c^2}, \label{eq:3dJ_TEe1}
\end{equation}
the $\tilde{q}_z$ integral converging at $\tilde{q}_z\sim{1}/\delta_\omega$, as expected.

\bibliography{retardation_paper.5}

\begin{thebibliography}{37}%
\makeatletter
\providecommand \@ifxundefined [1]{%
 \@ifx{#1\undefined}
}%
\providecommand \@ifnum [1]{%
 \ifnum #1\expandafter \@firstoftwo
 \else \expandafter \@secondoftwo
 \fi
}%
\providecommand \@ifx [1]{%
 \ifx #1\expandafter \@firstoftwo
 \else \expandafter \@secondoftwo
 \fi
}%
\providecommand \natexlab [1]{#1}%
\providecommand \enquote  [1]{``#1''}%
\providecommand \bibnamefont  [1]{#1}%
\providecommand \bibfnamefont [1]{#1}%
\providecommand \citenamefont [1]{#1}%
\providecommand \href@noop [0]{\@secondoftwo}%
\providecommand \href [0]{\begingroup \@sanitize@url \@href}%
\providecommand \@href[1]{\@@startlink{#1}\@@href}%
\providecommand \@@href[1]{\endgroup#1\@@endlink}%
\providecommand \@sanitize@url [0]{\catcode `\\12\catcode `\$12\catcode
  `\&12\catcode `\#12\catcode `\^12\catcode `\_12\catcode `\%12\relax}%
\providecommand \@@startlink[1]{}%
\providecommand \@@endlink[0]{}%
\providecommand \url  [0]{\begingroup\@sanitize@url \@url }%
\providecommand \@url [1]{\endgroup\@href {#1}{\urlprefix }}%
\providecommand \urlprefix  [0]{URL }%
\providecommand \Eprint [0]{\href }%
\providecommand \doibase [0]{http://dx.doi.org/}%
\providecommand \selectlanguage [0]{\@gobble}%
\providecommand \bibinfo  [0]{\@secondoftwo}%
\providecommand \bibfield  [0]{\@secondoftwo}%
\providecommand \translation [1]{[#1]}%
\providecommand \BibitemOpen [0]{}%
\providecommand \bibitemStop [0]{}%
\providecommand \bibitemNoStop [0]{.\EOS\space}%
\providecommand \EOS [0]{\spacefactor3000\relax}%
\providecommand \BibitemShut  [1]{\csname bibitem#1\endcsname}%
\let\auto@bib@innerbib\@empty
\bibitem [{\citenamefont {Rytov}(1953)}]{Rytov1953}%
  \BibitemOpen
  \bibfield  {author} {\bibinfo {author} {\bibfnamefont {S.~M.}\ \bibnamefont
  {Rytov}},\ }\href@noop {} {\emph {\bibinfo {title} {Theory of electric
  fluctuations and thermal radiation}}}\ (\bibinfo  {publisher} {Air Force
  Cambrige Research Center, Bedford, MA},\ \bibinfo {year} {1953})\BibitemShut
  {NoStop}%
\bibitem [{\citenamefont {Polder}\ and\ \citenamefont
  {Van~Hove}(1971)}]{Polder1971}%
  \BibitemOpen
  \bibfield  {author} {\bibinfo {author} {\bibfnamefont {D.}~\bibnamefont
  {Polder}}\ and\ \bibinfo {author} {\bibfnamefont {M.}~\bibnamefont
  {Van~Hove}},\ }\bibfield  {title} {\enquote {\bibinfo {title} {Theory of
  radiative heat transfer between closely spaced bodies},}\ }\href {\doibase
  10.1103/PhysRevB.4.3303} {\bibfield  {journal} {\bibinfo  {journal} {Phys.
  Rev. B}\ }\textbf {\bibinfo {volume} {4}},\ \bibinfo {pages} {3303--3314}
  (\bibinfo {year} {1971})}\BibitemShut {NoStop}%
\bibitem [{\citenamefont {Levin}\ \emph {et~al.}(1980)\citenamefont {Levin},
  \citenamefont {Polevoi},\ and\ \citenamefont {Rytov}}]{Levin1980}%
  \BibitemOpen
  \bibfield  {author} {\bibinfo {author} {\bibfnamefont {M.~L.}\ \bibnamefont
  {Levin}}, \bibinfo {author} {\bibfnamefont {V.~G.}\ \bibnamefont {Polevoi}},
  \ and\ \bibinfo {author} {\bibfnamefont {S.~M.}\ \bibnamefont {Rytov}},\
  }\bibfield  {title} {\enquote {\bibinfo {title} {Contribution to the theory
  of heat exchange due to a fluctuating electromagnetic field},}\ }\href@noop
  {} {\bibfield  {journal} {\bibinfo  {journal} {Sov. Phys. JETP}\ }\textbf
  {\bibinfo {volume} {52}},\ \bibinfo {pages} {1054} (\bibinfo {year}
  {1980})}\BibitemShut {NoStop}%
\bibitem [{\citenamefont {Loomis}\ and\ \citenamefont
  {Maris}(1994)}]{Loomis1994}%
  \BibitemOpen
  \bibfield  {author} {\bibinfo {author} {\bibfnamefont {Jackson~J.}\
  \bibnamefont {Loomis}}\ and\ \bibinfo {author} {\bibfnamefont {Humphrey~J.}\
  \bibnamefont {Maris}},\ }\bibfield  {title} {\enquote {\bibinfo {title}
  {Theory of heat transfer by evanescent electromagnetic waves},}\ }\href
  {\doibase 10.1103/PhysRevB.50.18517} {\bibfield  {journal} {\bibinfo
  {journal} {Phys. Rev. B}\ }\textbf {\bibinfo {volume} {50}},\ \bibinfo
  {pages} {18517--18524} (\bibinfo {year} {1994})}\BibitemShut {NoStop}%
\bibitem [{\citenamefont {Pendry}(1999)}]{Pendry1999}%
  \BibitemOpen
  \bibfield  {author} {\bibinfo {author} {\bibfnamefont {J.~B.}\ \bibnamefont
  {Pendry}},\ }\bibfield  {title} {\enquote {\bibinfo {title} {Radiative
  exchange of heat between nanostructures},}\ }\href {\doibase
  10.1088/0953-8984/11/35/301} {\bibfield  {journal} {\bibinfo  {journal}
  {Journal of Physics: Condensed Matter}\ }\textbf {\bibinfo {volume} {11}},\
  \bibinfo {pages} {6621--6633} (\bibinfo {year} {1999})}\BibitemShut {NoStop}%
\bibitem [{\citenamefont {Rytov}\ \emph {et~al.}(1989)\citenamefont {Rytov},
  \citenamefont {Kravtsov},\ and\ \citenamefont {Tatarskii}}]{Rytov1989}%
  \BibitemOpen
  \bibfield  {author} {\bibinfo {author} {\bibfnamefont {S.~M.}\ \bibnamefont
  {Rytov}}, \bibinfo {author} {\bibfnamefont {Y.~A.}\ \bibnamefont {Kravtsov}},
  \ and\ \bibinfo {author} {\bibfnamefont {V.~I.}\ \bibnamefont {Tatarskii}},\
  }\href@noop {} {\emph {\bibinfo {title} {Principles of statistical
  radiophysics}}}\ (\bibinfo  {publisher} {Springer-Verlag, Berlin
  Heidelberg},\ \bibinfo {year} {1989})\BibitemShut {NoStop}%
\bibitem [{\citenamefont {Joulain}\ \emph {et~al.}(2005)\citenamefont
  {Joulain}, \citenamefont {Mulet}, \citenamefont {Marquier}, \citenamefont
  {Carminati},\ and\ \citenamefont {Greffet}}]{Joulain2005}%
  \BibitemOpen
  \bibfield  {author} {\bibinfo {author} {\bibfnamefont {Karl}\ \bibnamefont
  {Joulain}}, \bibinfo {author} {\bibfnamefont {Jean-Philippe}\ \bibnamefont
  {Mulet}}, \bibinfo {author} {\bibfnamefont {Fran{\c c}ois}\ \bibnamefont
  {Marquier}}, \bibinfo {author} {\bibfnamefont {R{\'e}mi}\ \bibnamefont
  {Carminati}}, \ and\ \bibinfo {author} {\bibfnamefont {Jean-Jacques}\
  \bibnamefont {Greffet}},\ }\bibfield  {title} {\enquote {\bibinfo {title}
  {Surface electromagnetic waves thermally excited: Radiative heat transfer,
  coherence properties and casimir forces revisited in the near field},}\
  }\href {\doibase https://doi.org/10.1016/j.surfrep.2004.12.002} {\bibfield
  {journal} {\bibinfo  {journal} {Surface Science Reports}\ }\textbf {\bibinfo
  {volume} {57}},\ \bibinfo {pages} {59 -- 112} (\bibinfo {year}
  {2005})}\BibitemShut {NoStop}%
\bibitem [{\citenamefont {Volokitin}\ and\ \citenamefont
  {Persson}(2007)}]{Volokitin2007}%
  \BibitemOpen
  \bibfield  {author} {\bibinfo {author} {\bibfnamefont {A.~I.}\ \bibnamefont
  {Volokitin}}\ and\ \bibinfo {author} {\bibfnamefont {B.~N.~J.}\ \bibnamefont
  {Persson}},\ }\bibfield  {title} {\enquote {\bibinfo {title} {Near-field
  radiative heat transfer and noncontact friction},}\ }\href {\doibase
  10.1103/RevModPhys.79.1291} {\bibfield  {journal} {\bibinfo  {journal} {Rev.
  Mod. Phys.}\ }\textbf {\bibinfo {volume} {79}},\ \bibinfo {pages}
  {1291--1329} (\bibinfo {year} {2007})}\BibitemShut {NoStop}%
\bibitem [{\citenamefont {Song}\ \emph {et~al.}(2015)\citenamefont {Song},
  \citenamefont {Fiorino}, \citenamefont {Meyhofer},\ and\ \citenamefont
  {Reddy}}]{Song2015}%
  \BibitemOpen
  \bibfield  {author} {\bibinfo {author} {\bibfnamefont {Bai}\ \bibnamefont
  {Song}}, \bibinfo {author} {\bibfnamefont {Anthony}\ \bibnamefont {Fiorino}},
  \bibinfo {author} {\bibfnamefont {Edgar}\ \bibnamefont {Meyhofer}}, \ and\
  \bibinfo {author} {\bibfnamefont {Pramod}\ \bibnamefont {Reddy}},\ }\bibfield
   {title} {\enquote {\bibinfo {title} {Near-field radiative thermal transport:
  From theory to experiment},}\ }\href {\doibase 10.1063/1.4919048} {\bibfield
  {journal} {\bibinfo  {journal} {AIP Advances}\ }\textbf {\bibinfo {volume}
  {5}},\ \bibinfo {pages} {053503} (\bibinfo {year} {2015})}\BibitemShut
  {NoStop}%
\bibitem [{\citenamefont {Biehs}\ \emph {et~al.}(2020)\citenamefont {Biehs},
  \citenamefont {Messina}, \citenamefont {Venkataram}, \citenamefont
  {Rodriguez}, \citenamefont {Cuevas},\ and\ \citenamefont
  {Ben-Abdallah}}]{Biehs2020}%
  \BibitemOpen
  \bibfield  {author} {\bibinfo {author} {\bibfnamefont {S.-A.}\ \bibnamefont
  {Biehs}}, \bibinfo {author} {\bibfnamefont {R.}~\bibnamefont {Messina}},
  \bibinfo {author} {\bibfnamefont {P.~S.}\ \bibnamefont {Venkataram}},
  \bibinfo {author} {\bibfnamefont {A.~W.}\ \bibnamefont {Rodriguez}}, \bibinfo
  {author} {\bibfnamefont {J.~C.}\ \bibnamefont {Cuevas}}, \ and\ \bibinfo
  {author} {\bibfnamefont {B.}~\bibnamefont {Ben-Abdallah}},\ }\href@noop {}
  {\enquote {\bibinfo {title} {Near-field radiative heat transfer in many-body
  systems},}\ } (\bibinfo {year} {2020}),\ \bibinfo {note}
  {arXiv:2007.05604}\BibitemShut {NoStop}%
\bibitem [{\citenamefont {Chapuis}\ \emph
  {et~al.}(2008{\natexlab{a}})\citenamefont {Chapuis}, \citenamefont {Volz},
  \citenamefont {Henkel}, \citenamefont {Joulain},\ and\ \citenamefont
  {Greffet}}]{Chapuis2008}%
  \BibitemOpen
  \bibfield  {author} {\bibinfo {author} {\bibfnamefont {Pierre-Olivier}\
  \bibnamefont {Chapuis}}, \bibinfo {author} {\bibfnamefont {Sebastian}\
  \bibnamefont {Volz}}, \bibinfo {author} {\bibfnamefont {Carsten}\
  \bibnamefont {Henkel}}, \bibinfo {author} {\bibfnamefont {Karl}\ \bibnamefont
  {Joulain}}, \ and\ \bibinfo {author} {\bibfnamefont {Jean-Jacques}\
  \bibnamefont {Greffet}},\ }\bibfield  {title} {\enquote {\bibinfo {title}
  {Effects of spatial dispersion in near-field radiative heat transfer between
  two parallel metallic surfaces},}\ }\href {\doibase
  10.1103/PhysRevB.77.035431} {\bibfield  {journal} {\bibinfo  {journal} {Phys.
  Rev. B}\ }\textbf {\bibinfo {volume} {77}},\ \bibinfo {pages} {035431}
  (\bibinfo {year} {2008}{\natexlab{a}})}\BibitemShut {NoStop}%
\bibitem [{\citenamefont {Chapuis}\ \emph
  {et~al.}(2008{\natexlab{b}})\citenamefont {Chapuis}, \citenamefont {Laroche},
  \citenamefont {Volz},\ and\ \citenamefont {Greffet}}]{Chapuis2008-1}%
  \BibitemOpen
  \bibfield  {author} {\bibinfo {author} {\bibfnamefont {Pierre-Olivier}\
  \bibnamefont {Chapuis}}, \bibinfo {author} {\bibfnamefont {Marine}\
  \bibnamefont {Laroche}}, \bibinfo {author} {\bibfnamefont {Sebastian}\
  \bibnamefont {Volz}}, \ and\ \bibinfo {author} {\bibfnamefont {Jean-Jacques}\
  \bibnamefont {Greffet}},\ }\bibfield  {title} {\enquote {\bibinfo {title}
  {Near-field induction heating of metallic nanoparticles due to infrared
  magnetic dipole contribution},}\ }\href {\doibase 10.1103/PhysRevB.77.125402}
  {\bibfield  {journal} {\bibinfo  {journal} {Phys. Rev. B}\ }\textbf {\bibinfo
  {volume} {77}},\ \bibinfo {pages} {125402} (\bibinfo {year}
  {2008}{\natexlab{b}})}\BibitemShut {NoStop}%
\bibitem [{\citenamefont {Prunnila}\ and\ \citenamefont
  {Laakso}(2013)}]{Prunnila2013}%
  \BibitemOpen
  \bibfield  {author} {\bibinfo {author} {\bibfnamefont {Mika}\ \bibnamefont
  {Prunnila}}\ and\ \bibinfo {author} {\bibfnamefont {Sampo~J}\ \bibnamefont
  {Laakso}},\ }\bibfield  {title} {\enquote {\bibinfo {title} {Interlayer heat
  transfer in bilayer carrier systems},}\ }\href {\doibase
  10.1088/1367-2630/15/3/033043} {\bibfield  {journal} {\bibinfo  {journal}
  {New Journal of Physics}\ }\textbf {\bibinfo {volume} {15}},\ \bibinfo
  {pages} {033043} (\bibinfo {year} {2013})}\BibitemShut {NoStop}%
\bibitem [{\citenamefont {Mahan}(2017)}]{Mahan2017}%
  \BibitemOpen
  \bibfield  {author} {\bibinfo {author} {\bibfnamefont {G.~D.}\ \bibnamefont
  {Mahan}},\ }\bibfield  {title} {\enquote {\bibinfo {title} {Tunneling of heat
  between metals},}\ }\href {\doibase 10.1103/PhysRevB.95.115427} {\bibfield
  {journal} {\bibinfo  {journal} {Phys. Rev. B}\ }\textbf {\bibinfo {volume}
  {95}},\ \bibinfo {pages} {115427} (\bibinfo {year} {2017})}\BibitemShut
  {NoStop}%
\bibitem [{\citenamefont {Zhang}\ \emph {et~al.}(2018)\citenamefont {Zhang},
  \citenamefont {L\"u},\ and\ \citenamefont {Wang}}]{Zhang2018}%
  \BibitemOpen
  \bibfield  {author} {\bibinfo {author} {\bibfnamefont {Zu-Quan}\ \bibnamefont
  {Zhang}}, \bibinfo {author} {\bibfnamefont {Jing-Tao}\ \bibnamefont {L\"u}},
  \ and\ \bibinfo {author} {\bibfnamefont {Jian-Sheng}\ \bibnamefont {Wang}},\
  }\bibfield  {title} {\enquote {\bibinfo {title} {Energy transfer between two
  vacuum-gapped metal plates: Coulomb fluctuations and electron tunneling},}\
  }\href {\doibase 10.1103/PhysRevB.97.195450} {\bibfield  {journal} {\bibinfo
  {journal} {Phys. Rev. B}\ }\textbf {\bibinfo {volume} {97}},\ \bibinfo
  {pages} {195450} (\bibinfo {year} {2018})}\BibitemShut {NoStop}%
\bibitem [{\citenamefont {Wang}\ \emph {et~al.}(2018)\citenamefont {Wang},
  \citenamefont {Zhang},\ and\ \citenamefont {L\"u}}]{Wang2018}%
  \BibitemOpen
  \bibfield  {author} {\bibinfo {author} {\bibfnamefont {Jian-Sheng}\
  \bibnamefont {Wang}}, \bibinfo {author} {\bibfnamefont {Zu-Quan}\
  \bibnamefont {Zhang}}, \ and\ \bibinfo {author} {\bibfnamefont {Jing-Tao}\
  \bibnamefont {L\"u}},\ }\bibfield  {title} {\enquote {\bibinfo {title}
  {Coulomb-force-mediated heat transfer in the near field: Geometric effect},}\
  }\href {\doibase 10.1103/PhysRevE.98.012118} {\bibfield  {journal} {\bibinfo
  {journal} {Phys. Rev. E}\ }\textbf {\bibinfo {volume} {98}},\ \bibinfo
  {pages} {012118} (\bibinfo {year} {2018})}\BibitemShut {NoStop}%
\bibitem [{\citenamefont {Kamenev}(2018)}]{Kamenev2018}%
  \BibitemOpen
  \bibfield  {author} {\bibinfo {author} {\bibfnamefont {Alex}\ \bibnamefont
  {Kamenev}},\ }\href@noop {} {\enquote {\bibinfo {title} {Near-field heat
  transfer between disordered conductors},}\ } (\bibinfo {year} {2018}),\
  \bibinfo {note} {arXiv:1811.10187}\BibitemShut {NoStop}%
\bibitem [{\citenamefont {Wise}\ \emph {et~al.}(2020)\citenamefont {Wise},
  \citenamefont {Basko},\ and\ \citenamefont {Hekking}}]{Wise2020}%
  \BibitemOpen
  \bibfield  {author} {\bibinfo {author} {\bibfnamefont {Jonathan~L.}\
  \bibnamefont {Wise}}, \bibinfo {author} {\bibfnamefont {Denis~M.}\
  \bibnamefont {Basko}}, \ and\ \bibinfo {author} {\bibfnamefont {Frank W.~J.}\
  \bibnamefont {Hekking}},\ }\bibfield  {title} {\enquote {\bibinfo {title}
  {Role of disorder in plasmon-assisted near-field heat transfer between
  two-dimensional metals},}\ }\href {\doibase 10.1103/PhysRevB.101.205411}
  {\bibfield  {journal} {\bibinfo  {journal} {Phys. Rev. B}\ }\textbf {\bibinfo
  {volume} {101}},\ \bibinfo {pages} {205411} (\bibinfo {year}
  {2020})}\BibitemShut {NoStop}%
\bibitem [{\citenamefont {Ying}\ and\ \citenamefont
  {Kamenev}(2020)}]{Ying2020}%
  \BibitemOpen
  \bibfield  {author} {\bibinfo {author} {\bibfnamefont {Xuzhe}\ \bibnamefont
  {Ying}}\ and\ \bibinfo {author} {\bibfnamefont {Alex}\ \bibnamefont
  {Kamenev}},\ }\bibfield  {title} {\enquote {\bibinfo {title} {Plasmonic
  tuning of near-field heat transfer between graphene monolayers},}\ }\href
  {\doibase 10.1103/PhysRevB.102.195426} {\bibfield  {journal} {\bibinfo
  {journal} {Phys. Rev. B}\ }\textbf {\bibinfo {volume} {102}},\ \bibinfo
  {pages} {195426} (\bibinfo {year} {2020})}\BibitemShut {NoStop}%
\bibitem [{\citenamefont {Govorov}\ and\ \citenamefont
  {Chaplik}(1989)}]{Govorov1989}%
  \BibitemOpen
  \bibfield  {author} {\bibinfo {author} {\bibfnamefont {A.~O.}\ \bibnamefont
  {Govorov}}\ and\ \bibinfo {author} {\bibfnamefont {A.~V.}\ \bibnamefont
  {Chaplik}},\ }\bibfield  {title} {\enquote {\bibinfo {title} {Retardation
  effects in the relaxation of a two-dimensional electron plasma},}\
  }\href@noop {} {\bibfield  {journal} {\bibinfo  {journal} {Sov. Phys. JETP}\
  }\textbf {\bibinfo {volume} {68}},\ \bibinfo {pages} {1143} (\bibinfo {year}
  {1989})}\BibitemShut {NoStop}%
\bibitem [{\citenamefont {Fal'ko}\ and\ \citenamefont
  {Khmel'nitskii}(1989)}]{Falko1989}%
  \BibitemOpen
  \bibfield  {author} {\bibinfo {author} {\bibfnamefont {V.~I.}\ \bibnamefont
  {Fal'ko}}\ and\ \bibinfo {author} {\bibfnamefont {D.~I.}\ \bibnamefont
  {Khmel'nitskii}},\ }\bibfield  {title} {\enquote {\bibinfo {title} {What if a
  film conductivity exceeds the speed of light?}}\ }\href@noop {} {\bibfield
  {journal} {\bibinfo  {journal} {Sov. Phys. JETP}\ }\textbf {\bibinfo {volume}
  {68}},\ \bibinfo {pages} {1150} (\bibinfo {year} {1989})}\BibitemShut
  {NoStop}%
\bibitem [{\citenamefont {Volkov}\ and\ \citenamefont
  {Pavlov}(2014)}]{Volkov2014}%
  \BibitemOpen
  \bibfield  {author} {\bibinfo {author} {\bibfnamefont {V.~A.}\ \bibnamefont
  {Volkov}}\ and\ \bibinfo {author} {\bibfnamefont {V.~N.}\ \bibnamefont
  {Pavlov}},\ }\bibfield  {title} {\enquote {\bibinfo {title} {Radiative
  plasmon polaritons in multilayer structures with a two-dimensional electron
  gas},}\ }\href {\doibase 10.1134/S0021364014020118} {\bibfield  {journal}
  {\bibinfo  {journal} {JETP Letters}\ }\textbf {\bibinfo {volume} {99}},\
  \bibinfo {pages} {93--98} (\bibinfo {year} {2014})}\BibitemShut {NoStop}%
\bibitem [{\citenamefont {Muravev}\ \emph {et~al.}(2015)\citenamefont
  {Muravev}, \citenamefont {Gusikhin}, \citenamefont {Andreev},\ and\
  \citenamefont {Kukushkin}}]{Muravev2015}%
  \BibitemOpen
  \bibfield  {author} {\bibinfo {author} {\bibfnamefont {V.~M.}\ \bibnamefont
  {Muravev}}, \bibinfo {author} {\bibfnamefont {P.~A.}\ \bibnamefont
  {Gusikhin}}, \bibinfo {author} {\bibfnamefont {I.~V.}\ \bibnamefont
  {Andreev}}, \ and\ \bibinfo {author} {\bibfnamefont {I.~V.}\ \bibnamefont
  {Kukushkin}},\ }\bibfield  {title} {\enquote {\bibinfo {title} {Novel
  relativistic plasma excitations in a gated two-dimensional electron
  system},}\ }\href {\doibase 10.1103/PhysRevLett.114.106805} {\bibfield
  {journal} {\bibinfo  {journal} {Phys. Rev. Lett.}\ }\textbf {\bibinfo
  {volume} {114}},\ \bibinfo {pages} {106805} (\bibinfo {year}
  {2015})}\BibitemShut {NoStop}%
\bibitem [{\citenamefont {Gusikhin}\ \emph {et~al.}(2018)\citenamefont
  {Gusikhin}, \citenamefont {Muravev}, \citenamefont {Zagitova},\ and\
  \citenamefont {Kukushkin}}]{Gusikhin2018}%
  \BibitemOpen
  \bibfield  {author} {\bibinfo {author} {\bibfnamefont {P.~A.}\ \bibnamefont
  {Gusikhin}}, \bibinfo {author} {\bibfnamefont {V.~M.}\ \bibnamefont
  {Muravev}}, \bibinfo {author} {\bibfnamefont {A.~A.}\ \bibnamefont
  {Zagitova}}, \ and\ \bibinfo {author} {\bibfnamefont {I.~V.}\ \bibnamefont
  {Kukushkin}},\ }\bibfield  {title} {\enquote {\bibinfo {title} {Drastic
  reduction of plasmon damping in two-dimensional electron disks},}\ }\href
  {\doibase 10.1103/PhysRevLett.121.176804} {\bibfield  {journal} {\bibinfo
  {journal} {Phys. Rev. Lett.}\ }\textbf {\bibinfo {volume} {121}},\ \bibinfo
  {pages} {176804} (\bibinfo {year} {2018})}\BibitemShut {NoStop}%
\bibitem [{\citenamefont {Oriekhov}\ and\ \citenamefont
  {Levitov}(2020)}]{Oriekhov2020}%
  \BibitemOpen
  \bibfield  {author} {\bibinfo {author} {\bibfnamefont {D.~O.}\ \bibnamefont
  {Oriekhov}}\ and\ \bibinfo {author} {\bibfnamefont {L.~S.}\ \bibnamefont
  {Levitov}},\ }\bibfield  {title} {\enquote {\bibinfo {title} {Plasmon
  resonances and tachyon ghost modes in highly conducting sheets},}\ }\href
  {\doibase 10.1103/PhysRevB.101.245136} {\bibfield  {journal} {\bibinfo
  {journal} {Phys. Rev. B}\ }\textbf {\bibinfo {volume} {101}},\ \bibinfo
  {pages} {245136} (\bibinfo {year} {2020})}\BibitemShut {NoStop}%
\bibitem [{\citenamefont {Ordal}\ \emph {et~al.}(1985)\citenamefont {Ordal},
  \citenamefont {Bell}, \citenamefont {Alexander}, \citenamefont {Long},\ and\
  \citenamefont {Querry}}]{Ordal1985}%
  \BibitemOpen
  \bibfield  {author} {\bibinfo {author} {\bibfnamefont {M.~A.}\ \bibnamefont
  {Ordal}}, \bibinfo {author} {\bibfnamefont {Robert~J.}\ \bibnamefont {Bell}},
  \bibinfo {author} {\bibfnamefont {R.~W.}\ \bibnamefont {Alexander}}, \bibinfo
  {author} {\bibfnamefont {L.~L.}\ \bibnamefont {Long}}, \ and\ \bibinfo
  {author} {\bibfnamefont {M.~R.}\ \bibnamefont {Querry}},\ }\bibfield  {title}
  {\enquote {\bibinfo {title} {Optical properties of fourteen metals in the
  infrared and far infrared: Al, co, cu, au, fe, pb, mo, ni, pd, pt, ag, ti, v,
  and w.}}\ }\href {\doibase 10.1364/AO.24.004493} {\bibfield  {journal}
  {\bibinfo  {journal} {Appl. Opt.}\ }\textbf {\bibinfo {volume} {24}},\
  \bibinfo {pages} {4493--4499} (\bibinfo {year} {1985})}\BibitemShut {NoStop}%
\bibitem [{\citenamefont {Wang}\ \emph {et~al.}(2019)\citenamefont {Wang},
  \citenamefont {Bie}, \citenamefont {Cai}, \citenamefont {Ge}, \citenamefont
  {Ji}, \citenamefont {Jia}, \citenamefont {Gong}, \citenamefont {Zhang},
  \citenamefont {Wang},\ and\ \citenamefont {Xu}}]{Wang2019}%
  \BibitemOpen
  \bibfield  {author} {\bibinfo {author} {\bibfnamefont {Lei}\ \bibnamefont
  {Wang}}, \bibinfo {author} {\bibfnamefont {Mengli}\ \bibnamefont {Bie}},
  \bibinfo {author} {\bibfnamefont {Wei}\ \bibnamefont {Cai}}, \bibinfo
  {author} {\bibfnamefont {Lixin}\ \bibnamefont {Ge}}, \bibinfo {author}
  {\bibfnamefont {Zhichao}\ \bibnamefont {Ji}}, \bibinfo {author}
  {\bibfnamefont {Yonglei}\ \bibnamefont {Jia}}, \bibinfo {author}
  {\bibfnamefont {Ke}~\bibnamefont {Gong}}, \bibinfo {author} {\bibfnamefont
  {Xinzheng}\ \bibnamefont {Zhang}}, \bibinfo {author} {\bibfnamefont
  {Junqiao}\ \bibnamefont {Wang}}, \ and\ \bibinfo {author} {\bibfnamefont
  {Jingjun}\ \bibnamefont {Xu}},\ }\bibfield  {title} {\enquote {\bibinfo
  {title} {Giant near-field radiative heat transfer between ultrathin metallic
  films},}\ }\href {\doibase 10.1364/OE.27.036790} {\bibfield  {journal}
  {\bibinfo  {journal} {Opt. Express}\ }\textbf {\bibinfo {volume} {27}},\
  \bibinfo {pages} {36790--36798} (\bibinfo {year} {2019})}\BibitemShut
  {NoStop}%
\bibitem [{\citenamefont {Altshuler}\ and\ \citenamefont
  {Aronov}(1985)}]{Altshuler1985}%
  \BibitemOpen
  \bibfield  {author} {\bibinfo {author} {\bibfnamefont {B.~L.}\ \bibnamefont
  {Altshuler}}\ and\ \bibinfo {author} {\bibfnamefont {A.~G.}\ \bibnamefont
  {Aronov}},\ }\bibfield  {title} {\enquote {\bibinfo {title} {Chapter 1 -
  electron--electron interaction in disordered conductors},}\ }in\ \href
  {\doibase https://doi.org/10.1016/B978-0-444-86916-6.50007-7} {\emph
  {\bibinfo {booktitle} {Electron--Electron Interactions in Disordered
  Systems}}},\ \bibinfo {series} {Modern Problems in Condensed Matter
  Sciences}, Vol.~\bibinfo {volume} {10},\ \bibinfo {editor} {edited by\
  \bibinfo {editor} {\bibfnamefont {A.L.}\ \bibnamefont {Efros}}\ and\ \bibinfo
  {editor} {\bibfnamefont {M.}~\bibnamefont {Pollak}}}\ (\bibinfo  {publisher}
  {North-Holland, Amsterdam},\ \bibinfo {year} {1985})\ pp.\ \bibinfo {pages}
  {1 -- 153}\BibitemShut {NoStop}%
\bibitem [{\citenamefont {Lee}\ and\ \citenamefont
  {Ramakrishnan}(1985)}]{Lee1985}%
  \BibitemOpen
  \bibfield  {author} {\bibinfo {author} {\bibfnamefont {Patrick~A.}\
  \bibnamefont {Lee}}\ and\ \bibinfo {author} {\bibfnamefont {T.~V.}\
  \bibnamefont {Ramakrishnan}},\ }\bibfield  {title} {\enquote {\bibinfo
  {title} {Disordered electronic systems},}\ }\href {\doibase
  10.1103/RevModPhys.57.287} {\bibfield  {journal} {\bibinfo  {journal} {Rev.
  Mod. Phys.}\ }\textbf {\bibinfo {volume} {57}},\ \bibinfo {pages} {287--337}
  (\bibinfo {year} {1985})}\BibitemShut {NoStop}%
\bibitem [{\citenamefont {Kralik}\ \emph {et~al.}(2012)\citenamefont {Kralik},
  \citenamefont {Hanzelka}, \citenamefont {Zobac}, \citenamefont {Musilova},
  \citenamefont {Fort},\ and\ \citenamefont {Horak}}]{Kralik2012}%
  \BibitemOpen
  \bibfield  {author} {\bibinfo {author} {\bibfnamefont {Tomas}\ \bibnamefont
  {Kralik}}, \bibinfo {author} {\bibfnamefont {Pavel}\ \bibnamefont
  {Hanzelka}}, \bibinfo {author} {\bibfnamefont {Martin}\ \bibnamefont
  {Zobac}}, \bibinfo {author} {\bibfnamefont {Vera}\ \bibnamefont {Musilova}},
  \bibinfo {author} {\bibfnamefont {Tomas}\ \bibnamefont {Fort}}, \ and\
  \bibinfo {author} {\bibfnamefont {Michal}\ \bibnamefont {Horak}},\ }\bibfield
   {title} {\enquote {\bibinfo {title} {Strong near-field enhancement of
  radiative heat transfer between metallic surfaces},}\ }\href {\doibase
  10.1103/PhysRevLett.109.224302} {\bibfield  {journal} {\bibinfo  {journal}
  {Phys. Rev. Lett.}\ }\textbf {\bibinfo {volume} {109}},\ \bibinfo {pages}
  {224302} (\bibinfo {year} {2012})}\BibitemShut {NoStop}%
\bibitem [{\citenamefont {Yang}\ \emph {et~al.}(2018)\citenamefont {Yang},
  \citenamefont {Du}, \citenamefont {Su}, \citenamefont {Fu}, \citenamefont
  {Gong}, \citenamefont {He},\ and\ \citenamefont {Ma}}]{Yang2018}%
  \BibitemOpen
  \bibfield  {author} {\bibinfo {author} {\bibfnamefont {Jiang}\ \bibnamefont
  {Yang}}, \bibinfo {author} {\bibfnamefont {Wei}\ \bibnamefont {Du}}, \bibinfo
  {author} {\bibfnamefont {Yishu}\ \bibnamefont {Su}}, \bibinfo {author}
  {\bibfnamefont {Yang}\ \bibnamefont {Fu}}, \bibinfo {author} {\bibfnamefont
  {Shaoxiang}\ \bibnamefont {Gong}}, \bibinfo {author} {\bibfnamefont
  {Sailing}\ \bibnamefont {He}}, \ and\ \bibinfo {author} {\bibfnamefont
  {Yungui}\ \bibnamefont {Ma}},\ }\bibfield  {title} {\enquote {\bibinfo
  {title} {Observing of the super-planckian near-field thermal radiation
  between graphene sheets},}\ }\href {\doibase 10.1038/s41467-018-06163-8}
  {\bibfield  {journal} {\bibinfo  {journal} {Nature Communications}\ }\textbf
  {\bibinfo {volume} {9}},\ \bibinfo {pages} {4033} (\bibinfo {year}
  {2018})}\BibitemShut {NoStop}%
\bibitem [{\citenamefont {Hargreaves}(1969)}]{Hargreaves1969}%
  \BibitemOpen
  \bibfield  {author} {\bibinfo {author} {\bibfnamefont {C.~M.}\ \bibnamefont
  {Hargreaves}},\ }\bibfield  {title} {\enquote {\bibinfo {title} {Anomalous
  radiative transfer between closely-spaced bodies},}\ }\href {\doibase
  https://doi.org/10.1016/0375-9601(69)90264-3} {\bibfield  {journal} {\bibinfo
   {journal} {Physics Letters A}\ }\textbf {\bibinfo {volume} {30}},\ \bibinfo
  {pages} {491 -- 492} (\bibinfo {year} {1969})},\ \bibinfo {note} {more
  precise measurements were described in the Ph.~D. Thesis of C. M. Hargreaves
  (University of Leiden, 1973), reproduced in Ref.~\cite{Song2015}}\BibitemShut
  {NoStop}%
\bibitem [{\citenamefont {Song}\ \emph {et~al.}(2016)\citenamefont {Song},
  \citenamefont {Thompson}, \citenamefont {Fiorino}, \citenamefont {Ganjeh},
  \citenamefont {Reddy},\ and\ \citenamefont {Meyhofer}}]{Song2016}%
  \BibitemOpen
  \bibfield  {author} {\bibinfo {author} {\bibfnamefont {Bai}\ \bibnamefont
  {Song}}, \bibinfo {author} {\bibfnamefont {Dakotah}\ \bibnamefont
  {Thompson}}, \bibinfo {author} {\bibfnamefont {Anthony}\ \bibnamefont
  {Fiorino}}, \bibinfo {author} {\bibfnamefont {Yashar}\ \bibnamefont
  {Ganjeh}}, \bibinfo {author} {\bibfnamefont {Pramod}\ \bibnamefont {Reddy}},
  \ and\ \bibinfo {author} {\bibfnamefont {Edgar}\ \bibnamefont {Meyhofer}},\
  }\bibfield  {title} {\enquote {\bibinfo {title} {Radiative heat conductances
  between dielectric and metallic parallel plates with nanoscale gaps},}\
  }\href {\doibase 10.1038/nnano.2016.17} {\bibfield  {journal} {\bibinfo
  {journal} {Nature Nanotechnology}\ }\textbf {\bibinfo {volume} {11}},\
  \bibinfo {pages} {509--514} (\bibinfo {year} {2016})}\BibitemShut {NoStop}%
\bibitem [{\citenamefont {Sabbaghi}\ \emph {et~al.}(2020)\citenamefont
  {Sabbaghi}, \citenamefont {Long}, \citenamefont {Ying}, \citenamefont
  {Lambert}, \citenamefont {Taylor}, \citenamefont {Messner},\ and\
  \citenamefont {Wang}}]{Sabbaghi2020}%
  \BibitemOpen
  \bibfield  {author} {\bibinfo {author} {\bibfnamefont {Payam}\ \bibnamefont
  {Sabbaghi}}, \bibinfo {author} {\bibfnamefont {Linshuang}\ \bibnamefont
  {Long}}, \bibinfo {author} {\bibfnamefont {Xiaoyan}\ \bibnamefont {Ying}},
  \bibinfo {author} {\bibfnamefont {Lee}\ \bibnamefont {Lambert}}, \bibinfo
  {author} {\bibfnamefont {Sydney}\ \bibnamefont {Taylor}}, \bibinfo {author}
  {\bibfnamefont {Christian}\ \bibnamefont {Messner}}, \ and\ \bibinfo {author}
  {\bibfnamefont {Liping}\ \bibnamefont {Wang}},\ }\bibfield  {title} {\enquote
  {\bibinfo {title} {Super-planckian radiative heat transfer between macroscale
  metallic surfaces due to near-field and thin-film effects},}\ }\href
  {\doibase 10.1063/5.0008259} {\bibfield  {journal} {\bibinfo  {journal}
  {Journal of Applied Physics}\ }\textbf {\bibinfo {volume} {128}},\ \bibinfo
  {pages} {025305} (\bibinfo {year} {2020})}\BibitemShut {NoStop}%
\bibitem [{\citenamefont {Modest}(2013)}]{Modest2013}%
  \BibitemOpen
  \bibfield  {author} {\bibinfo {author} {\bibfnamefont {M.~F.}\ \bibnamefont
  {Modest}},\ }\href@noop {} {\emph {\bibinfo {title} {Radiative Heat
  Transfer}}}\ (\bibinfo  {publisher} {Academic Press, San Diego},\ \bibinfo
  {year} {2013})\BibitemShut {NoStop}%
\bibitem [{\citenamefont {Volokitin}\ and\ \citenamefont
  {Persson}(2001)}]{Volokitin2001}%
  \BibitemOpen
  \bibfield  {author} {\bibinfo {author} {\bibfnamefont {A.~I.}\ \bibnamefont
  {Volokitin}}\ and\ \bibinfo {author} {\bibfnamefont {B.~N.~J.}\ \bibnamefont
  {Persson}},\ }\bibfield  {title} {\enquote {\bibinfo {title} {Radiative heat
  transfer between nanostructures},}\ }\href {\doibase
  10.1103/PhysRevB.63.205404} {\bibfield  {journal} {\bibinfo  {journal} {Phys.
  Rev. B}\ }\textbf {\bibinfo {volume} {63}},\ \bibinfo {pages} {205404}
  (\bibinfo {year} {2001})}\BibitemShut {NoStop}%
\bibitem [{\citenamefont {Fu}\ and\ \citenamefont {Zhang}(2006)}]{Fu2006}%
  \BibitemOpen
  \bibfield  {author} {\bibinfo {author} {\bibfnamefont {C~J}\ \bibnamefont
  {Fu}}\ and\ \bibinfo {author} {\bibfnamefont {Z~M}\ \bibnamefont {Zhang}},\
  }\bibfield  {title} {\enquote {\bibinfo {title} {Nanoscale radiation heat
  transfer for silicon at different doping levels},}\ }\href@noop {} {\bibfield
   {journal} {\bibinfo  {journal} {International Journal of Heat and Mass
  Transfer}\ }\textbf {\bibinfo {volume} {49}},\ \bibinfo {pages} {1703--1718}
  (\bibinfo {year} {2006})}\BibitemShut {NoStop}%
\end{thebibliography}%

\end{document}